\documentclass{aastex7}

\newcommand\ucac{2MASS J09424023$-$4637176} 
\newcommand\mead{MEAD\,62}
\newcommand\meadb{MEAD\,62\,B}
\newcommand{\Mjup}{M$_{\mathrm{Jup}}$}
\newcommand{\Msun}{M$_{\mathrm{\odot}}$}
\newcommand{\Teff}{T$_{\mathrm{eff}}$}

\usepackage{hyperref}
\usepackage{cleveref}

\crefformat{section}{\S#2#1#3} 
\crefformat{subsection}{\S#2#1#3}
\crefformat{subsubsection}{\S#2#1#3}

\begin{document}

\title{The MIRI Excesses around Degenerates (MEAD) Survey I: A candidate cold brown dwarf in orbit around the nearby white dwarf \ucac}

\correspondingauthor{Lo\"ic Albert}
\email{loic.albert@umontreal.ca}

\author[0000-0003-0475-9375]{Lo\"ic Albert}
\affiliation{D\'epartement de physique and Institut Trottier de recherche sur les exoplan\`etes, Universit\'e de Montr\'eal, C.P. 6128, Succ. Centre-ville, Montr\'eal, H3C 3J7, Québec, Canada}
\affiliation{Observatoire du Mont-M\'egantic, Universit\'e de Montr\'eal, C.P. 6128, Succ. Centre-ville, Montr\'eal, H3C 3J7, Québec, Canada}
\email{loic.albert@umontreal.ca}

\author[0009-0008-7425-8609]{Sabrina Poulsen}
\affiliation{Homer L. Dodge Department of Physics and Astronomy, University of Oklahoma, 440 W. Brooks St., Norman, OK, 73019 USA}
\email{Sabrina.R.Poulsen-1@ou.edu}

\author[0000-0002-3307-1062]{\'Erika Le Bourdais}
\affiliation{D\'epartement de physique and Institut Trottier de recherche sur les exoplan\`etes, Universit\'e de Montr\'eal, C.P. 6128, Succ. Centre-ville, Montr\'eal, H3C 3J7, Québec, Canada}
\email{erika.le.bourdais@umontreal.ca}

\author[0000-0002-1783-8817]{John H. Debes}
\affiliation{The Space Telescope Science Institute, 3700 San Martin Dr., Baltimore, MD 21218, USA}
\affiliation{Aura, for ESA}
\email{debes@stsci.edu}

\author[0009-0003-5977-9581]{Samuel Boucher}
\affiliation{D\'epartement de physique and Institut Trottier de recherche sur les exoplan\`etes, Universit\'e de Montr\'eal, C.P. 6128, Succ. Centre-ville, Montr\'eal, H3C 3J7, Québec, Canada}
\email{samuel.boucher.3@umontreal.ca}

\author[0000-0001-6098-2235]{Mukremin Kilic}
\affiliation{Homer L. Dodge Department of Physics and Astronomy, University of Oklahoma, 440 W. Brooks St., Norman, OK, 73019 USA}
\email{kilic@ou.edu}

\author[0000-0001-8362-4094]{William Reach}
\affiliation{Space Science Institute, 4765 Walnut Street, Suite 205, Boulder, CO 80301, USA}
\email{wreach@spacescience.org}

\author[0000-0001-7106-4683]{Susan E. Mullally}
\affiliation{Space Telescope Science Institute, 3700 San Martin Dr, Baltimore, MD 21218, USA}
\email{smullally@stsci.edu}

\author[0000-0002-7698-3002]{Misty Cracraft}
\affiliation{Space Telescope Science Institute, 3700 San Martin Dr, Baltimore, MD 21218, USA}
\email{cracraft@stsci.edu}

\author[0009-0004-7656-2402]{Fergal Mullally}
\affiliation{Constellation, 1310 Point Street, Baltimore, MD 21231, USA}
\email{fergal.mullally@gmail.com}

\author[0000-0003-1863-4960]{Matthew De Furio}
\affiliation{Department of Astronomy, The University of Texas at Austin, 2515 Speedway, Stop C1400, Austin, TX 78712, USA}
\affiliation{NSF Astronomy and Astrophysics Postdoctoral Fellow}
\email{defurio@utexas.edu}

\author[0000-0001-5941-2286]{J. J. Hermes}
\affiliation{Department of Astronomy, Boston University, 725 Commonwealth Avenue, Boston, MA 02215, USA}
\email{jjhermes@bu.edu}

\author[0000-0003-0214-609X]{Scott J. Kenyon}
\affiliation{Smithsonian Astrophysical Observatory, 60 Garden Street, Cambridge, MA 02138, USA}
\email{kenyon@cfa.harvard.edu}

\author[0000-0001-9834-7579]{Carl Melis}
\affiliation{Department of Astronomy \& Astrophysics, University of California San Diego, La Jolla, CA 92093-0424, USA}
\email{cmelis@ucsd.edu}

\author[0000-0003-3786-3486]{Seth Redfield}
\affiliation{Astronomy Department and Van Vleck Observatory, Wesleyan University, Middletown, CT, USA}
\email{sredfield@wesleyan.edu}

\author[0000-0001-9064-5598]{M. C. Wyatt}
\affiliation{Institute of Astronomy, University of Cambridge, Madingley Road, Cambridge CB3 0HA, UK}
\email{wyatt@ast.cam.ac.uk}

%
%




\author[0000-0003-4609-4500]{Patrick Dufour}
\affiliation{D\'epartement de physique and Institut Trottier de recherche sur les exoplan\`etes, Universit\'e de Montr\'eal, C.P. 6128, Succ. Centre-ville, Montr\'eal, H3C 3J7, Québec, Canada}
\email{patrick.dufour@umontreal.ca}

\author[0009-0004-9728-3576]{David A. Golimowski}
\affiliation{Space Telescope Science Institute, 3700 San Martin Dr, Baltimore, MD 21218, USA}
\email{golim@stsci.edu}


\author[0009-0002-4970-3930]{Ashley Messier}
\affiliation{Department of Astronomy, Smith College, Northampton MA 01063, USA}
\email{amessie2@jh.edu}

\author[0000-0003-1748-602X]{Jay Farihi}
\affiliation{Department of Physics and Astronomy, University College London, London WC1E 6BT, UK}
\email{j.farihi@ucl.ac.uk}






\begin{abstract}

The MIRI Excesses Around Degenerates Survey is a Cycle 2 James Webb Space Telescope (JWST) Survey program designed to image nearby white dwarfs in the mid-IR with the MIRI imaging mode. Only a handful of white dwarfs have previously been observed beyond 8~\micron. This survey gathered observations for 56 white dwarfs within 25~pc at 10 and 15~\micron, probing each white dwarf for unresolved IR excesses, IR flux deficits indicative of collision induced absorption, or resolved substellar companions. We present in this paper observations of our first target, \ucac\, (also UCAC4 217-039132), henceforth called \mead. It is a magnetic DA white dwarf with an estimated age of $7.6^{+1.7}_{-2.2}$\,Gyr. A red candidate companion, \meadb, about 2 magnitudes fainter at 15\,$\mu$m than the white dwarf is detected at an apparent separation of 1\farcs95. If confirmed, \meadb\, would be a $0.014^{+0.002}_{-0.003}$\,\Msun\, brown dwarf with T$_{\rm eff} = 343^{+7}_{-11}$\,K, according to ATMO2020 evolutionary models. Although its red F1000W$-$F1500W color is similar to background galaxies, \meadb\, is consistent with being an unresolved point-source from empirical PSF fitting. A false positive analysis yields an expectation number of 0.66 red (F1000W$-$F1500$ \geq +0.80$\,mag) unresolved sources within the same separation (r$\leq2$ arcsec) for the entire MEAD survey. Thus, this candidate companion as likely to be an actual companion as a false-positive unresolved background galaxy. Additional observations to measure common proper motion or sample the SED are warranted to confirm the nature of \meadb. A deep near-infrared imaging detection is achievable from the ground while JWST is needed at longer infrared wavelengths.


\end{abstract}



\section{Introduction} \label{sec:intro}

The development of infrared detector technologies has elevated white dwarfs to prime targets in the search for ultracool objects — such as substellar companions or cool debris disks — whose spectral energy distributions (SEDs) peak at wavelengths longer than 1 micron. One advantage is that white dwarf SEDs can be modeled accurately, enabling detection of excess infrared emission from a unseen cool companions at higher confidence. Also, their low luminosity offers better companion-to-star contrasts for direct imaging techniques of Gyr old systems. Historically, these advantages were quickly recognized and impacted numerous fields. Using the first infrared photometers to observe white dwarfs, the initial mass function could be studied down to substellar masses \citep{probst.1982,probst.1983}, the first brown dwarf, GD~165B, was identified as a companion \citep{becklin.1988,zuckerman.1992} and debris disks were found around G29-38 and GD~362 which opened a unique avenue to studying planet formation and evolution \citep{zuckerman.1987, becklin.2005, kilic.2005}. Despite the importance of white dwarfs, there remains vast gaps in our knowledge surrounding them, including the occurrence rates of substellar companions or dust \citep{barber2012, rocchetto15, farihi16, debes11}, as well as the behavior of collision-induced absorption (CIA) in their photosphere at longer wavelengths \citep{Blouin2017, blouin.2024}.
The launch of the JWST ushers in an exciting new era of white dwarf studies. The incredible gain in sensitivity and spatial resolution at wavelengths between 8-20~\micron\ in particular now allow nearby white dwarfs to be probed for old (Gyr), self-luminous exoplanet companions - characterized by effective temperatures of $\leq200$\,K and masses comparable to Jupiter’s - having survived the red giant/asymptotic giant evolutionary phases \citep{poulsen.2024}. Recently, four white dwarfs were targeted in a pilot survey using deep MIRI imaging with two candidate brown dwarfs identified around two of the white dwarfs \citep{mullally.2024}. Also, the transiting Jupiter-sized planet WD\,1856+534\,b could be detected in emission \citep{limbach.2025}.

JWST’s unparalleled mid-infrared sensitivity also enables the detection of brown dwarf companions around nearby white dwarfs, potentially as cool as Jupiter - offering rare benchmarks for studying the complex atmospheres of ultracool worlds. However, even though M dwarf companions to WDs are common, brown dwarfs are not: \citet{farihi.2005} estimate a brown dwarf companion frequency of $<0.5$\% based on a search around 261 WDs. Despite three decades of searching, only 11 close, detached white dwarf-brown dwarf binaries have been found with orbital periods ranging from 68 min \citep{Casewell18} to $\approx10$ hours \citep[][and references therein]{French24}. Wide white dwarf-brown dwarf system provide ideal benchmarks to test the brown dwarf evolutionary models since they do not go through the common envelope evolution like the close binaries. Resolved white dwarf-brown dwarf systems can be identified through common proper motion, and we currently know 7 such systems with separations ranging from 69 to $\sim20~000$AU and with spectral types from L1 to Y1 \citep{French23}. WD~0806$-$661B \citep{luhman.2011} has the latest spectral type (Y1) in this sample and has an estimated $T_{\rm eff}=$ 325-350 K \citep{leggett.2017}.

The MIRI Excesses Around Degenerates (MEAD) Survey is a Cycle 2 Survey program designed to do a volume limited survey of Gaia-confirmed white dwarfs within 25~pc in the F1000W and F1500W filters of the MIRI imager. The Survey was designed to take advantage of scheduling gaps in JWST operations, and so each white dwarf is observed with shallow imaging that is nevertheless sensitive to exoplanets and brown dwarfs. In general, giant planets or low-mass brown dwarf companions are expected to be ultracool (T$_{\rm eff} \leq 300$\,K) and appear very red (F1000W-F1500W $>>1$\,mag) based on exoplanet formation \citep{mordasini.2012, linder.2019} and BD evolutionary models \citep{phillips.2020, marley.2021, burrows.1997}. But so few objects of this temperature are known, let alone have photometry in the mid-IR, that the range of color that they span remains to be established observationally. Only Jupiter, Eps\,Ind\,Ab and a few other objects have published mid-IR photometry (See section~\ref{sec:cmd}).

We report here observations of the very first MEAD target, \ucac, which surprisingly showed a close candidate companion within 2\arcsec. In this paper, we include a short description of the MEAD survey (Sec. \ref{sec:survey}), describe the observations and data reduction (Sec. \ref{sec:obsdatared}) and photometric extraction (Sec. \ref{sec:candidateselection}). We then investigate whether the candidate is a viable substellar object by comparing its photometry to known brown dwarfs and cold planets in color-magnitudes diagrams (Sec. \ref{sec:cmd}), and by checking if its point-spread function is consistent with a point-source (Sec. \ref{sec:resolvedextended}). Assuming it is a real companion, we place limits on its spectral type (Sec. \ref{sec:spt}), and we model the spectral energy distribution of the white dwarf host (Sec. \ref{sec:whitedwarf}) to constrain the brown dwarf age, mass and temperature (Sec. \ref{sec:parameters}). We perform a false-positive analysis by counting red point-sources in the MEAD survey (Sec. \ref{sec:falsepositives}) before discussing our findings (Sec. \ref{sec:discussion}) and concluding (Sec. \ref{sec:conclusion}).

\section{MEAD Survey Strategy}\label{sec:survey}
 
We selected all $d<25$\,pc candidates from the Gentile-Fusillo {\em Gaia} EDR3 WD candidates catalog with a WD probability $>$95\%, $>15\sigma$ significant parallax, and \Teff\ = 3500-25000 K assuming a pure H composition. This resulted in 249 individual WDs. Spectroscopic follow-up of {\em Gaia} WDs with $d<40$~pc shows that more than 99\% of the candidates are indeed WDs \citep{Tremblay20,OBrien2023}, and $>$85\% of our sample has already been spectroscopically characterized \citep{Hollands2018,Tremblay20,McCleery2018}. We then cross-matched the predicted positions of the WDs in 2010.5589 (the average epoch of the ALLWISE survey) with a search radius of 3$^{\prime\prime}$. We removed 20 sources where a main sequence companion was closer than $\sim$10$^{\prime\prime}$ in order to avoid serious detector saturation from the primary star. The remaining 229 objects are well distributed across the sky, covering \Teff\ from 3900~K to 22000~K and estimated masses between 0.4~\Msun\ and 1.14~\Msun. The only bias to this survey is sky coverage, due to the scheduling constraints of the JWST Observatory. Short visits ($\leq100$\,minutes) for 229 targets were defined in our GO/Cycle 2 JWST \emph{Survey} program (PID 3964, PI Poulsen), of which 56 were carried out to fill schedule gaps.


Our observations had a goal of obtaining 10~\micron\ photometry with at least SNR of 50 for our faintest targets, as well as a faint point source sensitivity of 4.5~$\mu$Jy ($>$5$\sigma$) in both filters. For reference, these sensitivities are within a factor of 10 (for F1000W) and 5 (for F1500W) compared to the sensitivities achieved in six pointings of the deep galaxy CEERS program \citep{yang23}. Our primary science goal is to observe each WD with sufficient precision to detect $>$9\% (3$\sigma$) excesses or decrements in both filters. Another science goal is to detect faint resolved candidate substellar objects for further common proper motion follow-up. This survey is sensitive to bound companions with fluxes $>$2.7-4.5~$\mu$Jy, sensitive enough to detect planets and brown dwarfs in wide orbits. More details about the survey are outside the scope of this paper and are reported in a companion paper (S. Poulsen, submitted).

\section{Observations and Data Reduction}\label{sec:obsdatared}

On 2024 February 8th, during visit 62 of program 3964, \ucac\, was observed, henceforth referred to as \mead, which is a DA white dwarf at 20.47 pc \citep{Bailer21}. Its celestial coordinates at the time of the MIRI observation is $\alpha_{2000} = 9$:42:40.40, $\delta_{2000} = -$46:37:18.7 at epoch 2024.107. As all MEAD observations, it consisted in imaging sequences through two filters, F1000W and F1500W, at four dither positions. The total integration time was 222\,s in both filters, using  a single integration of N$_{\rm group}=20$ per dither in the FASTR1 detector readout mode.

For each set of observations, the individual dithered images were processed with build 10.0 of the JWST calibration pipeline. They were run with mostly default parameters through pipelines calwebb\_detector1 and calwebb\_image2. The only custom parameter used was setting the rejection threshold for the jump step in calwebb\_detector1 to 5.0. After calwebb\_image2 is run, the individual \*cal.fits images are stacked and a median sky image is created (per pixel), and is then subtracted from each cal image to create a set of median sky (background) subtracted images. This subtraction removes the background as well as any remaining detector effects. These background subtracted images are then combined in the next stage of the pipeline.
 
In calwebb\_image3, the stage in the pipeline that resamples and combines the different dithers, the following parameters were set. In the tweakreg step, the images were compared to GAIADR3 as a reference catalog, specifying that only three targets needed to match to GAIA to be considered a match. In the resample step, the ‘gaussian’ resample kernel was used and the weight\_type was set to ‘exptime’. The outlier detection step used ‘scale’ values that are double the default, or ‘1.0 0.8’. The scale parameters control the derivative used to identify bad pixels. There are two values because the first value applies to detecting the primary cosmic ray, and the second value is used for masking lower level bad pixels associated with the primary. The final output image is a resampled version of the combined dithered, background subtracted images. For each filter, there is one output image, with sky levels approximately zero.

\section{Photometry and Candidate Companion Selection}\label{sec:candidateselection}

\subsection{Photometry}\label{sec:photometry}

We ran Source Extractor \citep{bertin.1996} on the resampled stack of both F1000W and F1500W filters. We used a high threshold of detection of 2 contiguous pixels with 3 sigma above background noise level ({\tt DETECT\_THRESH = 3} and {\tt DETECT\_MINAREA = 2}) and the default parameters otherwise. We set {\tt WEIGHT\_MAP = VAR\_MAP} and built the weight map from the {\tt VAR\_POISSON} fits extension for each stack. We adopted the ``windowed'' outputs for the PSF position and shape (e.g. {\tt XWIN\_IMAGE, YWIN\_IMAGE, FWHM\_IMAGE}) and aperture extraction for the photometry (e.g. {\tt FLUX\_APER}). We used the {\tt jwst\_miri\_apcorr\_0010.fits} definitions for the aperture radius of 4.48/5.72\,pixels and calibrated to absolute fluxes using aperture corrections of 1.45/1.50 for the F1000W/F1500W filters. This resulted in catalogues containing 255 and 125 sources in the F1000W and F1500W images, respectively. For sources appearing in only one band, we estimated the flux upper limit in the other band by measuring the flux in randomly positioned apertures across the image to calculate an average scatter (1$\sigma$), and finally adopting the 3$\sigma$ level as our upper limit.





\subsection{Candidate Companion} \label{sec:candidate}



\begin{figure}
    \centering
    \includegraphics[width=0.5\textwidth]{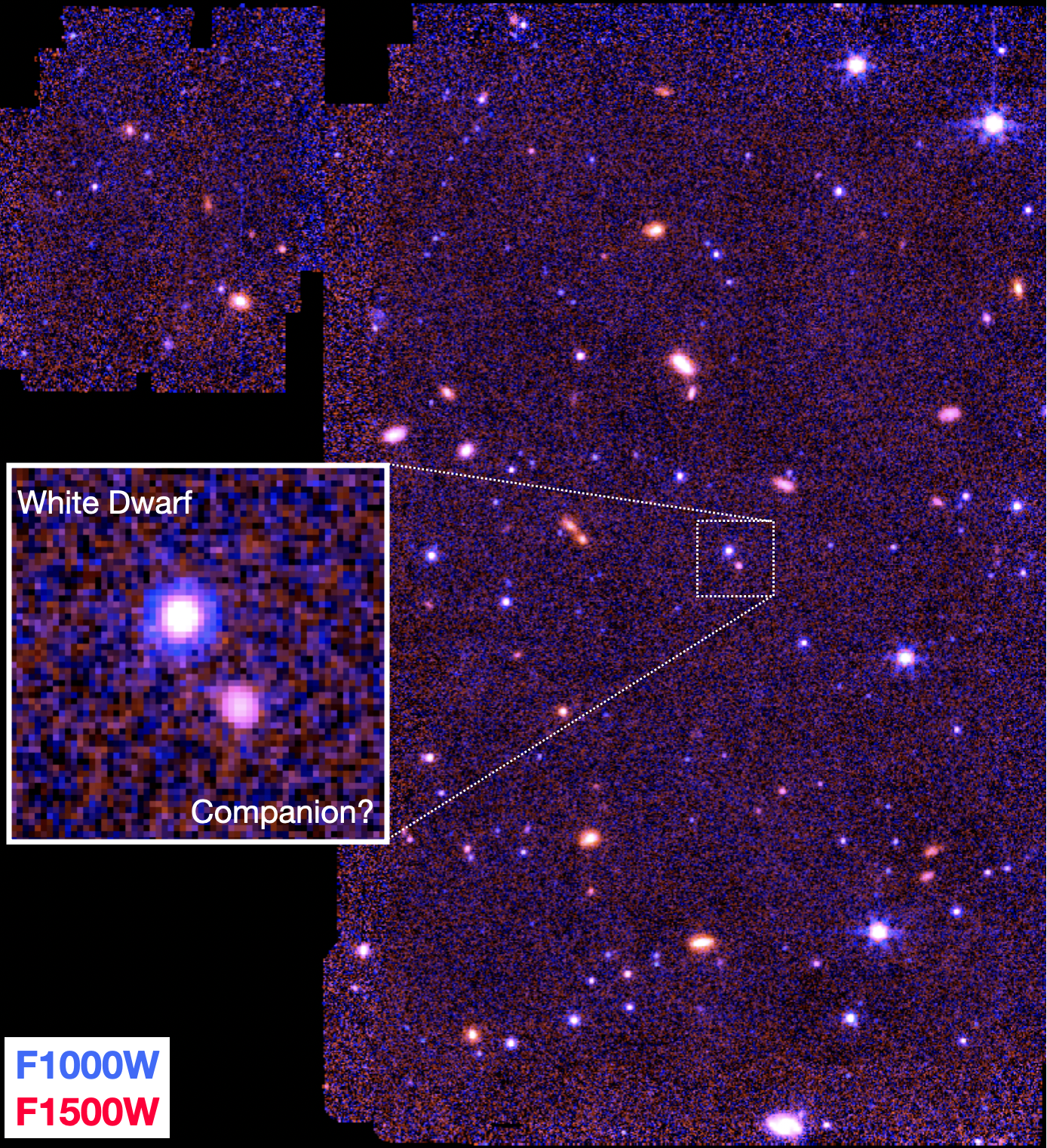}
    \caption{Color composite of the \mead~MIRI field with the F1000W band coded in blue and the F1500W coded in red showing the color diversity of sources in the field. A candidate companion is found 1.9\,\arcsec away from the white dwarf.}
    \label{fig:colorimage}
\end{figure}

Fig.~\ref{fig:colorimage} shows a color composite image of the MIRI field for \mead. 
A candidate companion, \meadb, with a projected separation of $1.953\pm0.002$\,\arcsec\, (40.0\,au) at a celestial position angle of $213.0\pm0.2$\,$^\circ$\,was quickly identified upon image inspection, in both bands. With Vega magnitudes of F1000W=$16.62\pm0.02$\,mag and F1500W = $14.94\pm0.02$\,mag, it has a very red colour of F1000W$-$F1500W = $1.67\pm0.03$\,mag. It is about 2 magnitudes fainter than the central WD which has F1000W=$14.24\pm0.01$\,mag and F1500W = $14.25\pm0.01$\,mag with a neutral color of F1000W$-$F1500W = $0.00\pm0.01$\,mag.

\section{Nature of the Candidate Companion}\label{sec:nature}

\subsection{Color-Magnitude Diagrams}\label{sec:cmd}

\begin{figure}
    \centering
    \includegraphics[width=1.0\linewidth]{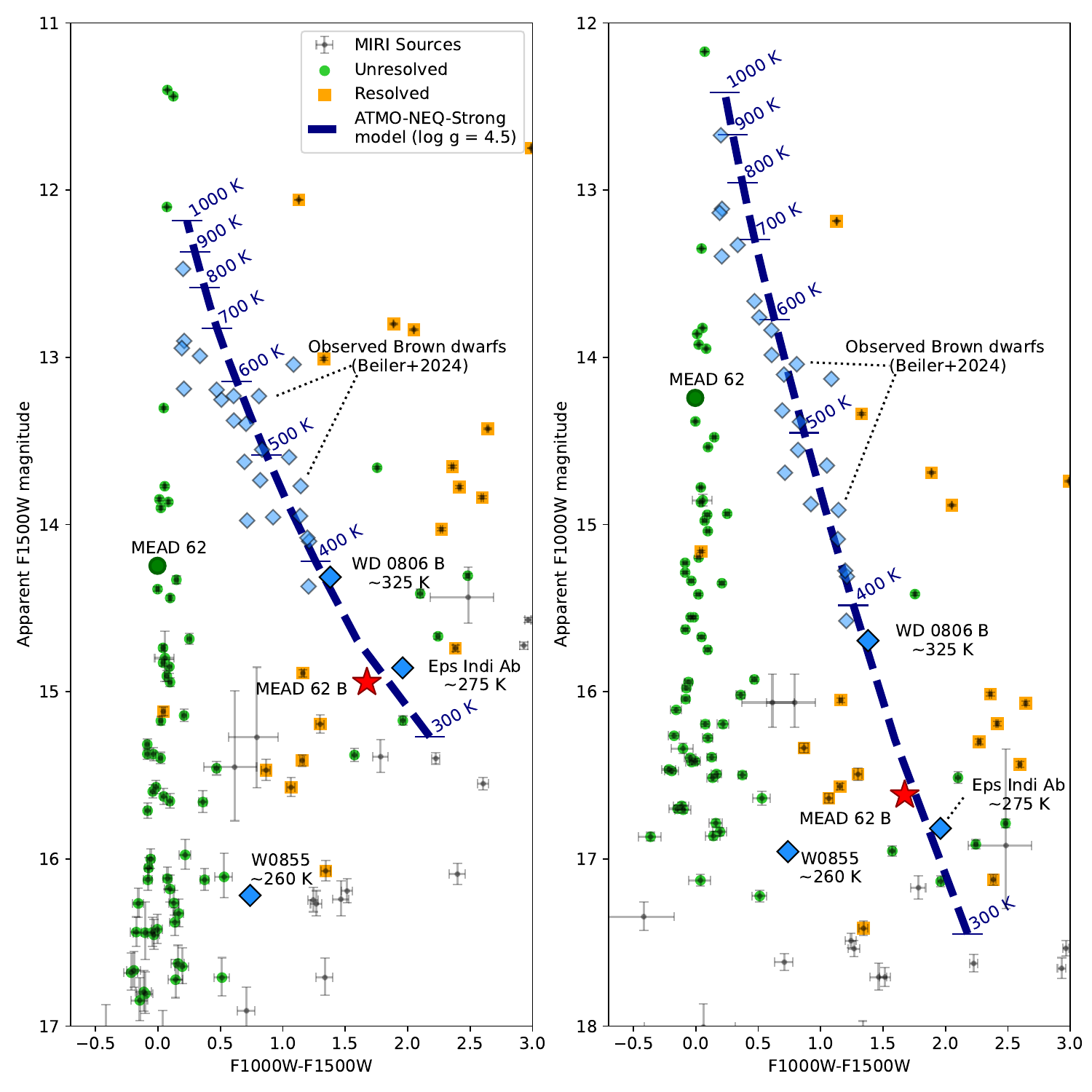}
    \caption{Apparent F1500W (left) and F1000W (right) magnitudes versus F1000W-F1500W colors for all sources detected in both filters at $\geq10$-$\sigma$ in the MIRI field of view for \mead. \meadb\, is consistent with a $\sim300$\,K\, brown dwarf when compared to an ATMO-2020 evolutionary model in these color-magnitude diagrams. 
    Individual catalogue sources are in black. The resolved sources (galaxies) are tagged as orange squares while unresolved point-sources are assigned green circles. The \mead\, white dwarf is shown as a large green circle while the candidate companion, \meadb, is the large red star symbol. The ATMO2020 evolutionary model track for $\log g=4.5$ with strong chemical disequilibrium is overlaid as a navy blue dashed line for substellar objects ranging between 1000\,K and 300\,K \citep{phillips.2020}. In addition, actual MIRI measurements of brown dwarfs in the same temperature range are plotted as pale blue diamonds and confirm that the ATMO2020 model can be trusted, especially at T$_{\rm eff}\leq 600$\,K \citep{beiler.2024}. Plotted as large blue diamonds are WD\,$0806-661$\,B, the sole Y-type brown dwarf associated to a white dwarf \citep{voyer.2025}; W0855, the coldest known brown dwarf \citep{luhman.2014,leggett.2021}; and the $6.3\pm0.6$\,\Mjup\, exoplanet Eps Ind Ab \citep{matthews.2024}. They represent the coldest objects having measured F1000W and F1500W magnitudes.}
    \label{fig:cmd}
\end{figure}

To assess if the companion may be physically associated with \mead, we plot the apparent F1000W (left) and F1500W (right) magnitudes vs. F1000W-F1500W color in Figure\,\ref{fig:cmd}. In these color-magnitude diagrams (CMDs), we plot all sources detected in both MIRI filters with error bars. Furthermore, the resolved sources (galaxies) are highlighted with orange squares while the unresolved point-sources are highlighted as plain green circles. Definition of the two categories is based on the FWHM and signal-to-noise ratio (SNR) in the F1000W band using the following boundary conditions for point sources as well as resolved sources: $25\leq$SNR$\leq50$ \& FWHM$\leq4.3$\,pixels; $50\leq$SNR$\leq100$ \& FWHM$\leq3.5$\,pixels; SNR$\geq100$ \& FWHM$\leq3.4$\,pixels. Sources having SNR$\leq25$ are not categorized. 

Clearly, unresolved sources concentrate on a vertical sequence with neutral $F1000W-F1500W \approx 0$ color and this is where the target white dwarf lies (large green circle). Galaxies are spread over the diagram with redder $0.7\leq F1000W - F1500W \leq 2.5$ color. Uncategorized sources tend to occupy the same color space as galaxies, hinting that their PSFs may be unresolved too. Also, a handful of sources categorized as unresolved have red colors indicating they may be intrinsically red or, more likely, wrongly categorized. One of these is our companion candidate, \meadb\, (large red star symbol). To help interpret the nature of this candidate, an ATMO+2020 brown dwarf model is overplotted as a thick dark blue dashed line \citep{phillips.2020}. The model with strong non-equilibrium chemistry and a $\log{g}=4.5$ is drawn for a range of effective temperatures between 1000\,K and 300\,K. Additionaly, 23 brown dwarfs observed with MIRI/LRS have their synthetic photometry overplotted as pale blue diamonds \citep{beiler.2024}. They follow rather well the model sequence with a spread of roughly 0.5\,mag in color. Three more known ultracool objects are overplotted as large blue diamonds in these CMDs: WD\,$0806-661$\,B, a brown dwarf companion with $T_{\rm eff}\sim325$\,K at extremely large separation ($\sim2500$\,au) from a white dwarf (WD\,$0806-661$) \citep{luhman.2011,voyer.2025}; Eps\,Ind\,Ab, a jupiter-mass planet with $T_{\rm eff}\sim275$\,K orbiting the nearby system Eps\,Indi\,A \citep{matthews.2024}; and WISE J085510.83$-$071442.5 (hereafter W0855), the coldest known brown dwarf with $T_{\rm eff}\sim260$\,K\citep{luhman.2014,leggett.2021}. The F1000W photometry of WD\,0806\,B was synthesized from the default MIRI/LRS spectrum ({\tt extract1d.fits}) found in MAST as part of observation 3 of program ID 1276. While the F1500W photometry comes from the F1500W imaging observation 4 of the same program. Absolute magnitudes are F1000W =$14.14\pm0.02$, F1500W=$12.76\pm0.01$. Similarly, the W0855 absolute photometry is F1000W =$15.399\pm0.006$, F1500W=$14.661\pm0.003$, synthesized from the MIRI/MRS spectrum of program ID 1230. 

Our candidate companion lands very close to the model sequence in between its 300\,K and 400\,K markers, also very close to the Eps\,Ind\,Ab giant planet. Based on this model track, our candidate should also be significantly cooler than WD\,$0806-661$\,B. However, the coldest brown dwarf, W0855, is positioned more than 1 magnitude off the model track and the Eps\,Ind\,Ab exoplanet of similar temperature. This is a testament to the complexity in modeling ultracool atmospheres (e.g. see \citet{leggett.2021, albert.2025,kuhnle.2025}) and a motivation to search for similar objects such as our \meadb\, candidate.

\subsection{Unresolved or Extended?}\label{sec:resolvedextended}

A low-mass substellar or planetary-mass companion such as our candidate is expected to harbor very red F1000W-F1500W$\geq0.5$ colors when detected in the mid-infrared but absent from near-infrared or optical surveys. Alternatively, galaxies can display similar red colors in those mid-infrared wavebands. Analysis of the  candidate's point-spread function (PSF) is crucial in discarding our candidate if it is resolved or retaining it as a candidate if it is not.

As evidenced in the color image (Figure\,\ref{fig:colorimage}) and in the color-magnitude diagram (Figure\,\ref{fig:cmd}), most objects with red colors ($0.5 \leq $ F1000W-F1500W $\leq 2.5$) are resolved, extended galaxies. They also are relatively bright (F1000W$\leq16$, F1500W$\leq17$) thus likely nearby. Experimenting with various galaxy templates available in the JWST exposure time calculator sheds light on the type of galaxy spectral features best explaining these colors. First, the population of galaxies whose PAH emission lines at $\simeq8\mu$m and $\simeq11\mu$m are redshifted in and out of the 2 MIRI bandpasses, can explain colors of $0.5 \leq $ F1500W-F1500W$ \leq 1.5$. But so can star-forming galaxies in the nearby universe ($z \leq 0.1$) whose cold dust emission peaking at $\approx60\mu$m extends down to 15$\mu$m. Interestingly, only star-forming galaxies can produce the reddest colors, F1500W-F1500W$ \geq 1.5$. This is important as it confirms that the main false positives of our survey are nearby galaxies, thus more easily resolved than distant ones.



A first assessment at establishing our candidate companion nature comes from the Source Extractor measurements of {\tt FWHM\_IMAGE} and {\tt ELONGATION} for all catalogued sources. The FWHM of the companion, FWHM=3.37/4.68 pixels, is consistent with point sources in the field, FWHM=$3.22\pm0.16$/$4.73\pm0.29$, in the F1000W/F1500W band. The elongation is the major-to-minor axis ratio after an ellipse is fit to the shape of the PSF. The companion has an elongation of 1.083/1.111 while points sources have $1.10\pm0.07$/$1.08\pm0.06$. It is worth noting that these Source Extractor measurements are performed on {\tt i2d.fits} stacks from only 4 dither positions such that the resulting PSF shape can be dependent on the actual image resampling. Nevertheless, these measurements are consistent with the companion being an unresolved point-source.

\begin{figure}[h!]
    \centering
    \includegraphics[width=0.49\linewidth]{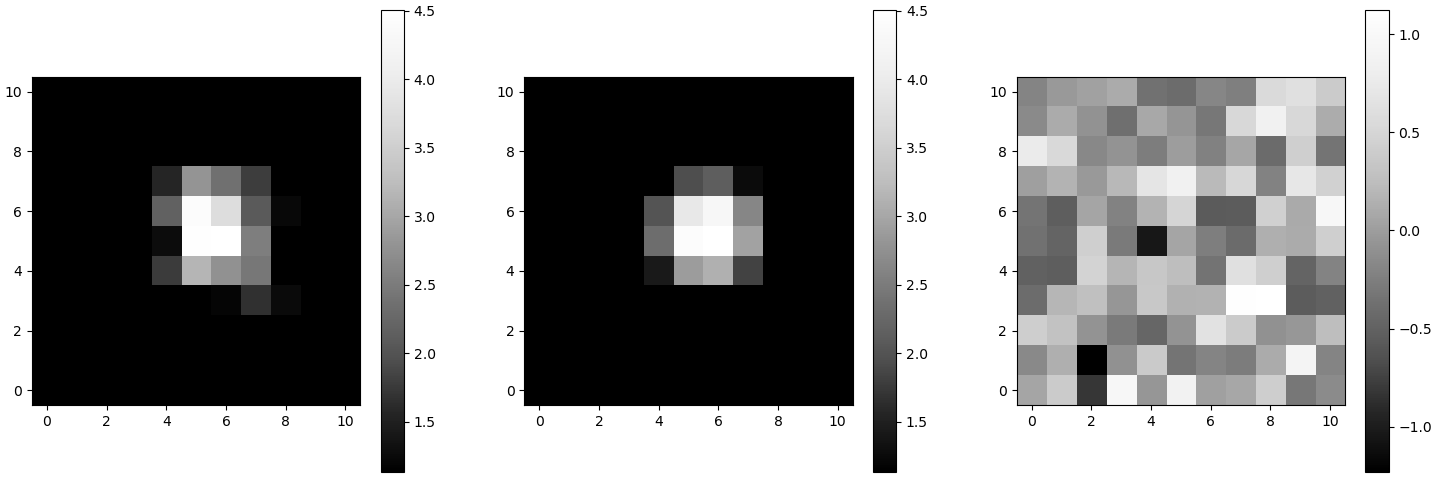}
    \includegraphics[width=0.49\linewidth]{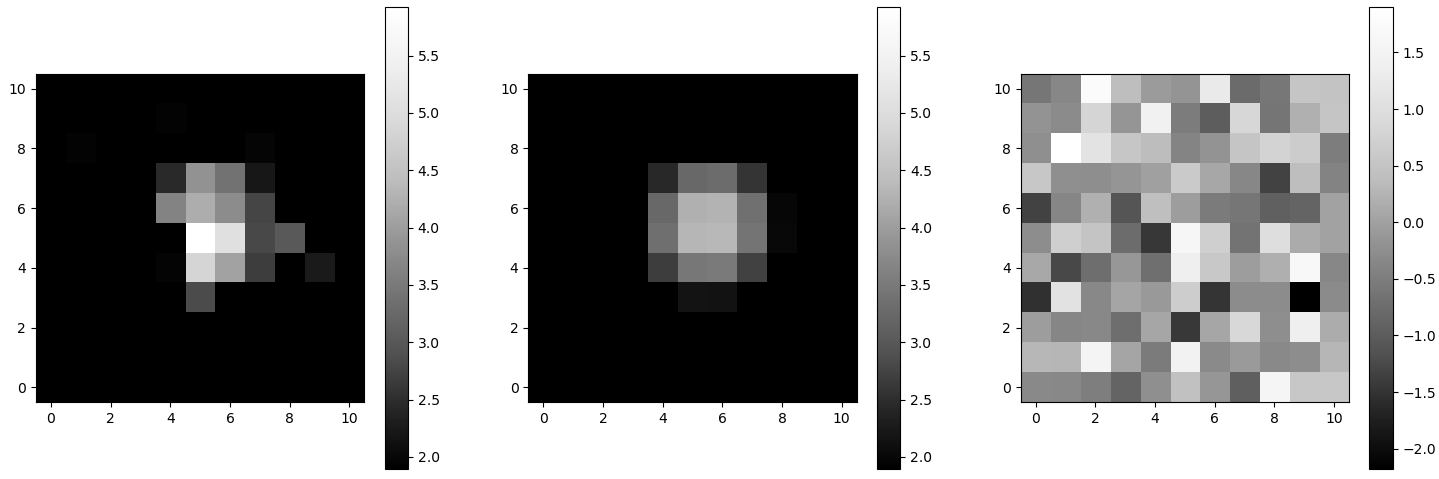}
    \includegraphics[width=0.49\linewidth]{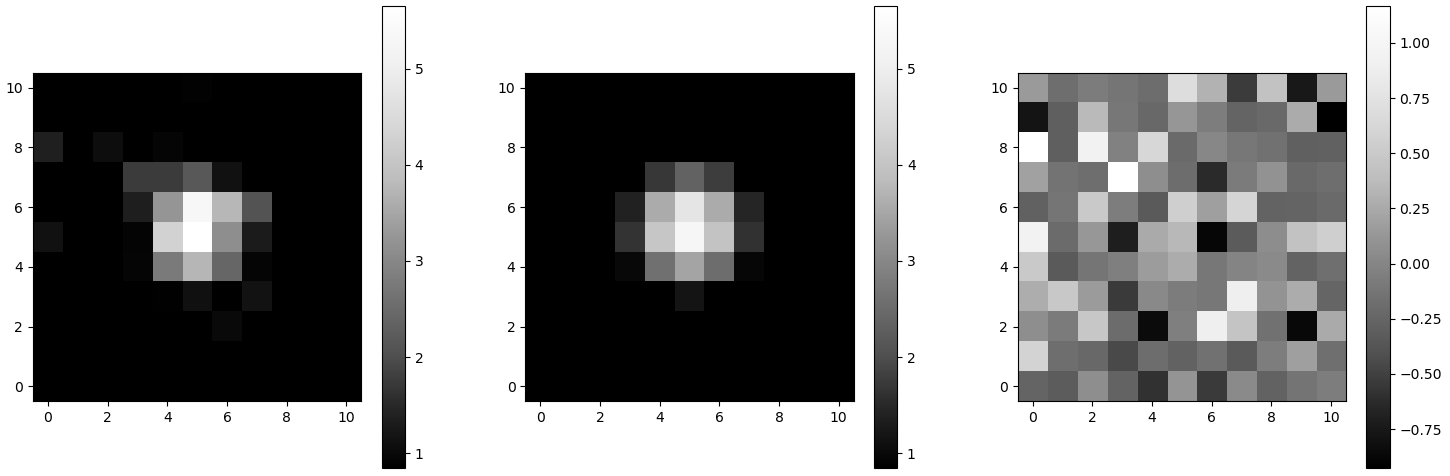}
    \includegraphics[width=0.49\linewidth]{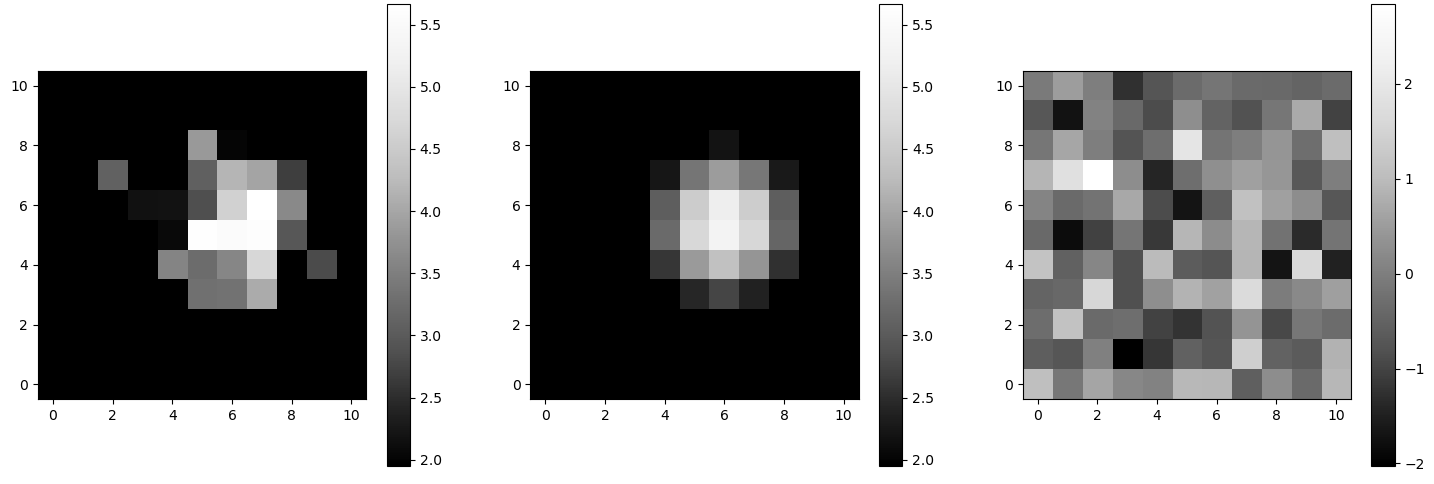}
    \includegraphics[width=0.49\linewidth]{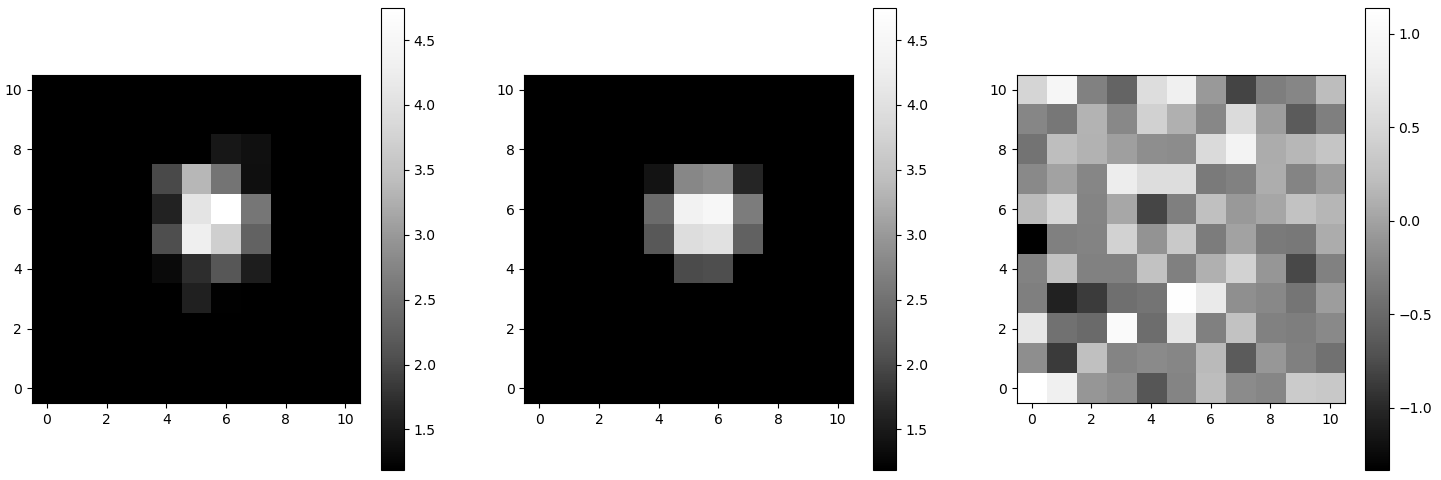}
    \includegraphics[width=0.49\linewidth]{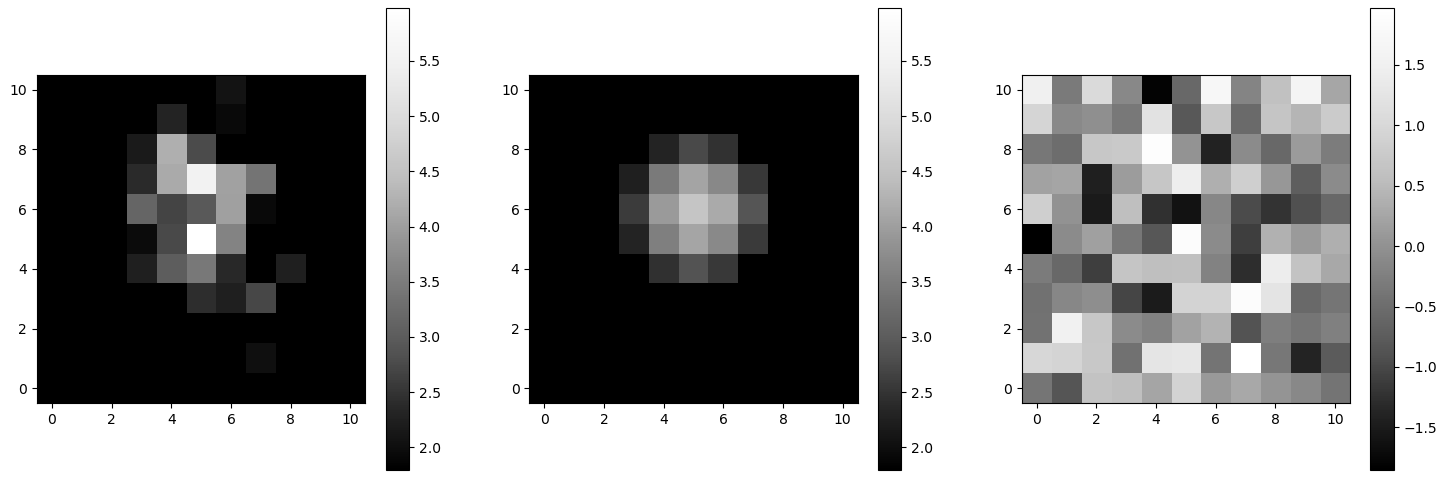}
    \includegraphics[width=0.49\linewidth]{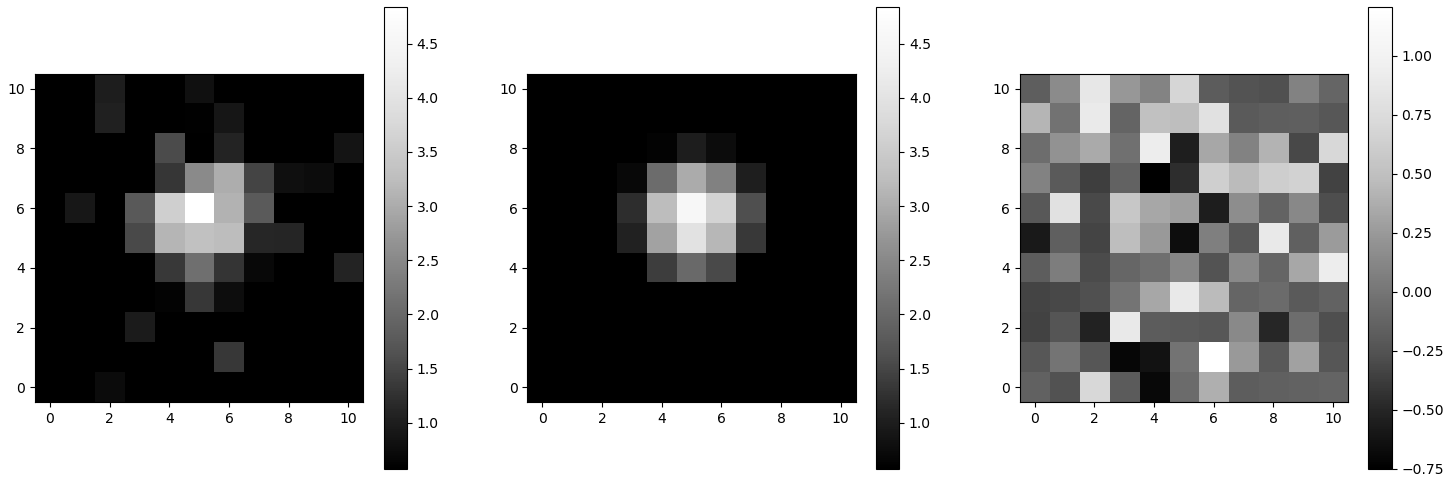}
    \includegraphics[width=0.49\linewidth]{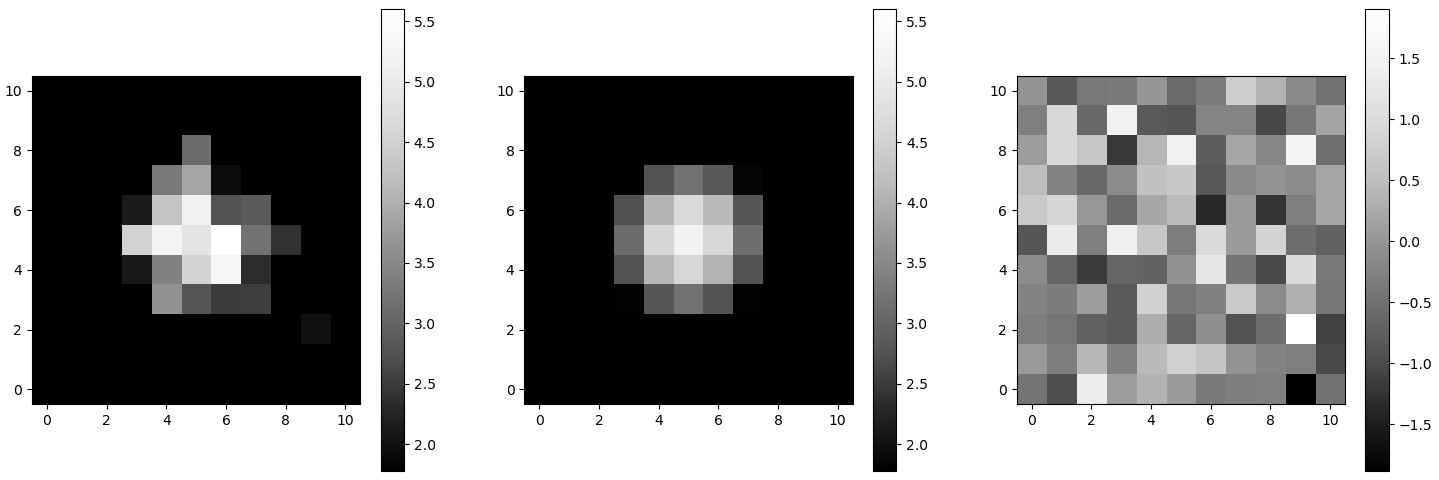}
    \caption{Empirical point-spread function fitting indicates that \meadb\, is an unresolved point-source. State of the art model PSFs \citep{libralato.2024} were used to fit the individual exposure for each of the 4 dither positions, assuming a single point source. The left panel is for the F1000W images while the right panel is for F1500W images. For each row of each panel, the observation is on the left, the best-fit PSF model, in the middle, and the \emph{observation $-$ model} residuals, on the right. No apparent flux excess is seen in the residuals map which is consistent with the obtained range of ePSF fitting reduced chi-squares: 0.99-1.27 (F1000W) and 1.00-1.49 (F1500W).}
    \label{fig:epsf}
\end{figure}

Our second approach to assess the nature of our candidate companion consists in performing a careful empirical point-spread function (ePSF) fitting analysis as described in \cite{defurio.2023}. That method makes use of well characterized and oversampled model PSFs empirically constructed for each specific MIRI filter, available at 25 detector positions (5$\times$5 grid) \citep{libralato.2024}. The ePSF method registers, resamples and scales those model PSFs to fit the observations at each individual dither position. We assumed a single point-source model and recovered the best-fit amplitude, x/y detector position and posterior distribution of these in a Monte-Carlo-Markov-Chain (MCMC) framework. The reduced chi-squared returned for the four F1000W dithers span $0.99 \leq \chi^2_\nu \leq 1.27$, consistent with the point-source hypothesis. For the four F1500W dithers, chi-squared span $1.00 \leq \chi^2_\nu \leq 1.49$, also consistent with a point-source. In Figure\,\ref{fig:epsf}, the observation, best-fit PSF model and subtraction residuals are plotted for each filter and dither position. It shows how the PSF changes due to the sampling at different sub-pixel detector positions. Also, the residual maps do not display apparent excesses above the noise, suggesting that the candidate companion PSF is indeed unresolved.

Based on these analyses, the companion seems to be an unresolved point-source, and thus we retain it as a candidate.

\subsection{\meadb\, spectral type constraints} \label{sec:spt}
%
%

While MIRI observations suggest the candidate to be an object with $T_{\rm eff}\leq350$\,K, a detection at shorter wavelengths could disprove the companion hypothesis. Gaia DR3, 2MASS and WISE surveyed that field but none detects a companion object to \mead. However, brown dwarfs of that temperature harbor a 4-5$\mu$m peak. Given its on-sky separation, this companion would be unresolved in the WISE survey thus, in principle, could be detected as an infrared excess. However, the measured W1 and W2 magnitudes of \mead~are consistent with a Rayleigh-Jeans slope, meaning no flux excess is observed. Lets consider the distance to \mead\, of 20.47~pc and adopt the brown dwarfs absolute magnitude versus spectral type calibration of \citet{dupuy.2017} and $K-W1$ and $K-W2$ colors tabulated by \citet[Table 1]{skrzypek.2015}. Then, any BD of spectral type earlier than roughly T2 (T7) would have produced a W1 (W2) excess, at the $1\sigma$ level (the spectral type - magnitude relation driving this dispersion, $\sigma=0.5$\,mag). At a higher confidence level, 3$\sigma$, the lack of a W2 excess strongly supports a spectral type of $\geq$T0. To summarize, the WISE non-detection constrains any companion to be -- most likely (1$\sigma$) -- of spectral type later than T7 ($T_{\rm eff} \leq 700$~K), or -- certainly (3$\sigma$) -- later than T0 ($T_{\rm eff} \leq 1000$~K). That is perfectly consistent with the 250-350\,K temperature range suggested by models shown in Figure\,\ref{fig:cmd}.

\subsection{White dwarf modeling}\label{sec:whitedwarf}

We constructed a spectral energy distribution (SED) for the white dwarf primary, \mead, using available photometry from Gaia DR3 ($B_p$, $R_p$ and $G$), 2MASS ($J$, $H$ and $K$) and WISE ($W1$ and $W2$). \mead\ is a magnetic DA white dwarf \citep{OBrien2023}. We used the photometric method to fit the spectral energy distribution of \mead\ using pure H
atmosphere models \citep{Bergeron2019}. The best fitting model has $T_{\rm eff} = 5968\pm124$K and $\log{g} = 8.02\pm0.04$ (cm s$^{-2}$), which are nearly identical to the values obtained by \citet{Obrien2024}. We note that the temperature of \mead, although close, remains above the $\simeq5000$\,K threshold for when an ill-modeled collision-induced absorption starts altering the SED around $2.4~\mu$m (see \cite{blouin.2024} and references therein). 

Constraints on the total age of the white dwarf + brown dwarf system were obtained using the \emph{wdwarfdate} Python package of \cite{kiman.2022}. Assuming a hydrogen-rich atmosphere, it uses the temperature and surface gravity to establish the white dwarf final mass of $0.60\pm0.02$\,\Msun. Then, it uses the initial-to-final mass relation of \cite{cummings.2018} to estimate the progenitor main sequence mass of $1.35^{+0.40}_{-0.25}$\,\Msun and the lifetime on the main sequence of $4.12^{+4.24}_{-2.31}$\,Gyr. Finally, that is added to the white dwarf cooling age of $2.51^{+0.29}_{-0.20}$\,Gyr based on the \citet{Bedard20} evolutionary models to constrain the total age to $6.64^{+4.15}_{-2.14}$\,Gyr. Uncertainties are given at the 16th and 84th percentiles.

\begin{figure}
    \centering
    \includegraphics[width=0.8\linewidth]{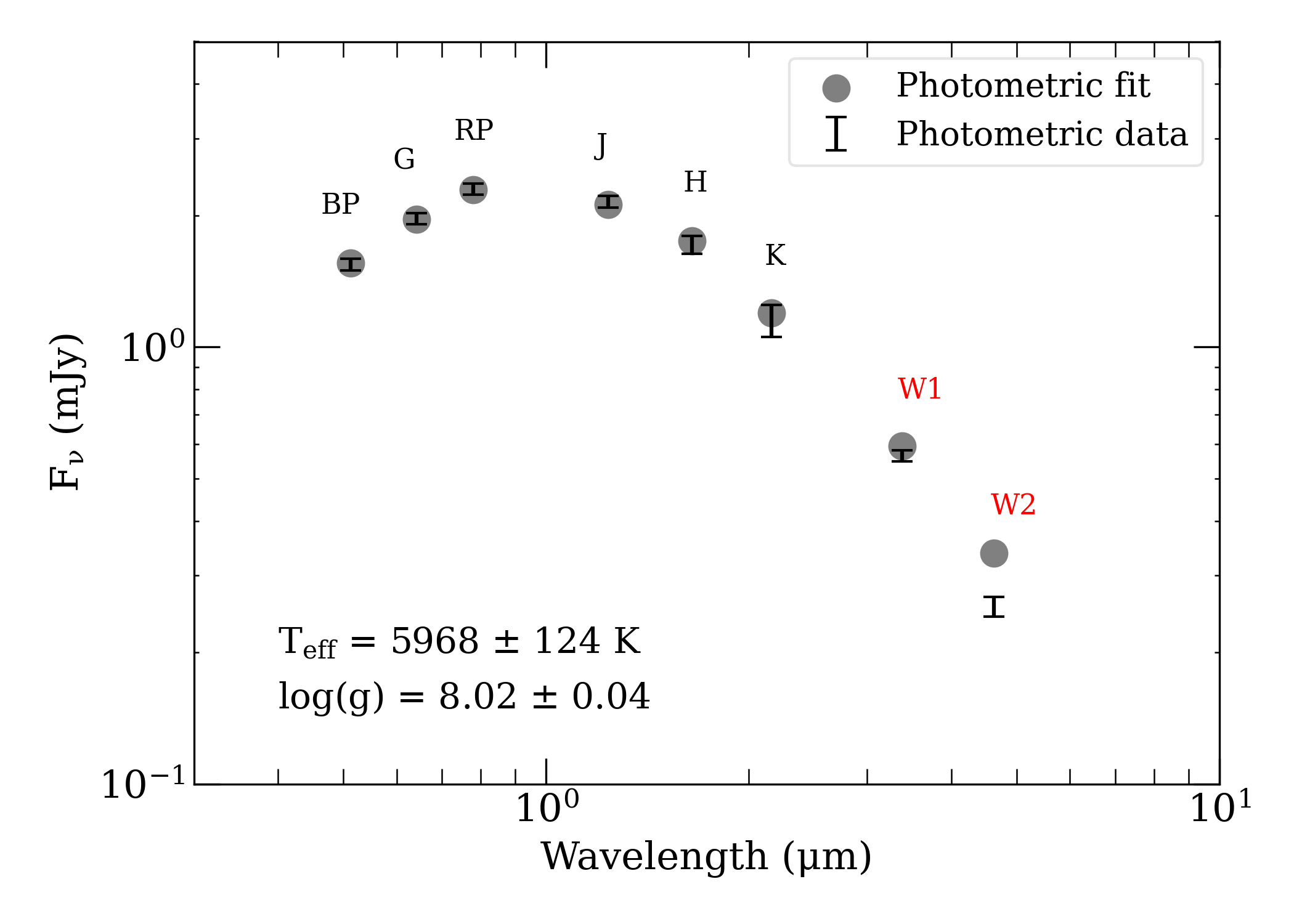}
    \includegraphics[width=1.0\linewidth]{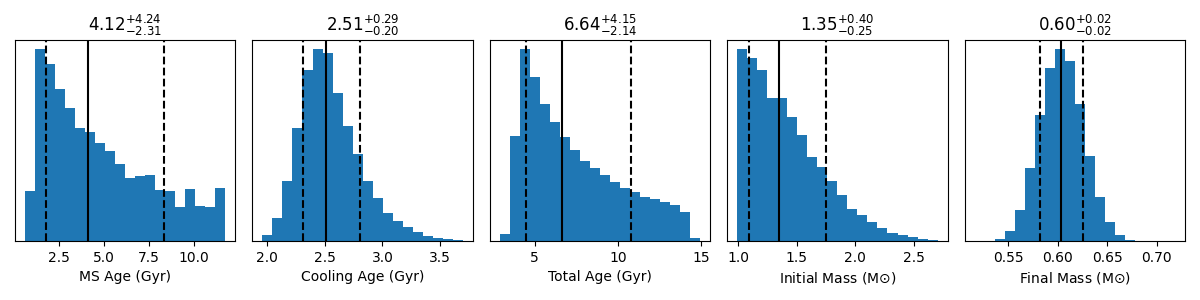}
    \caption{White dwarf physical properties. Top) Fit of the white dwarf spectral energy distribution probed using 6 photometric bands (W1 and W2 excluded from the fit) converges to a DA white dwarf with $T_{\rm eff} = 5968\pm124$K and surface gravity of $\log g = 8.02\pm0.04$. The fit has $\chi^2=0.75$. Bottom) Best estimates of the final white dwarf mass, initial main sequence mass, white dwarf cooling age, main sequence lifetime and total system age based on the \emph{wdwarfdate} Python package \citep{kiman.2022} using as inputs $T_{\rm eff} = 5968\pm124$\,K and $\log g=8.02\pm0.04$. }
    \label{fig:wdproperties}
\end{figure}

\subsection{Physical parameters of \meadb}\label{sec:parameters}

Assuming that brown dwarf evolutionary models can be trusted, the mass and temperature of the companion can be constrained given its measured apparent magnitudes, F1000W=$16.69\pm0.02$ and F1500W=$14.99\pm0.02$, known Gaia DR3 distance, and estimated age assuming coevality with the WD, $6.6^{+4.2}_{-2.1}$\,Gyr. We used the ATMO2020 evolutionary models \citep{phillips.2020} with strong non-equilibrium chemistry as they are relatively successful at reproducing Y-dwarf SEDs and they conveniently tabulate MIRI filter magnitudes. One limitation is that they are only defined for ages of $\leq10$\,Gyr. We ran this analysis using a Monte Carlo Markov Chain (MCMC) approach to retrieve a posterior distribution for the mass and age as well as an error scaling factor. The forward model interpolated the ATMO2020 grid at given ages and masses to yield F1000W and F1500W magnitudes. We used a uniform prior for the mass between 0.005 and 0.075\,\Msun\,and the white dwarf age distribution given by \emph{wdwarfdate} as the prior for the age, capped at 10\,Gyr.

\begin{figure}[!h]
    \centering
    \includegraphics[width=0.41\linewidth]{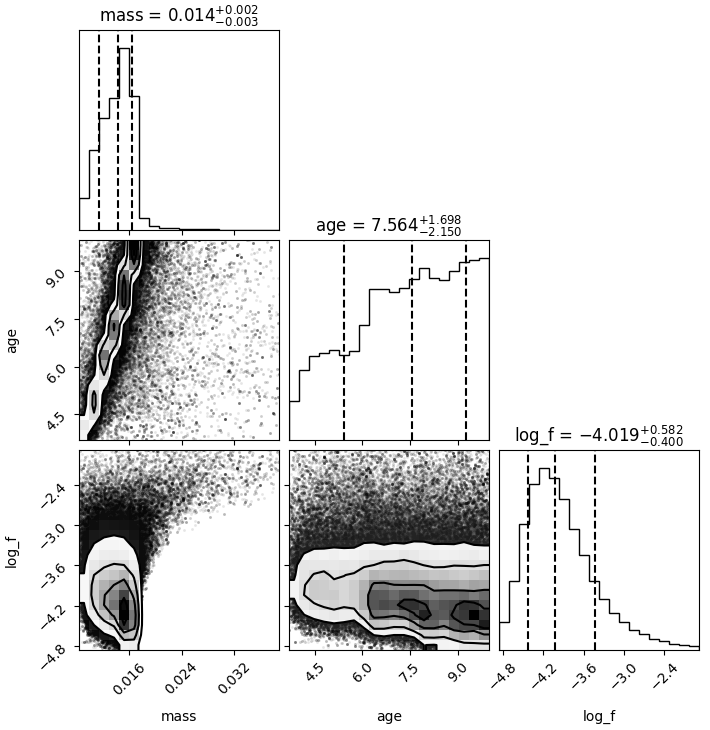}
    \includegraphics[width=0.55\linewidth]{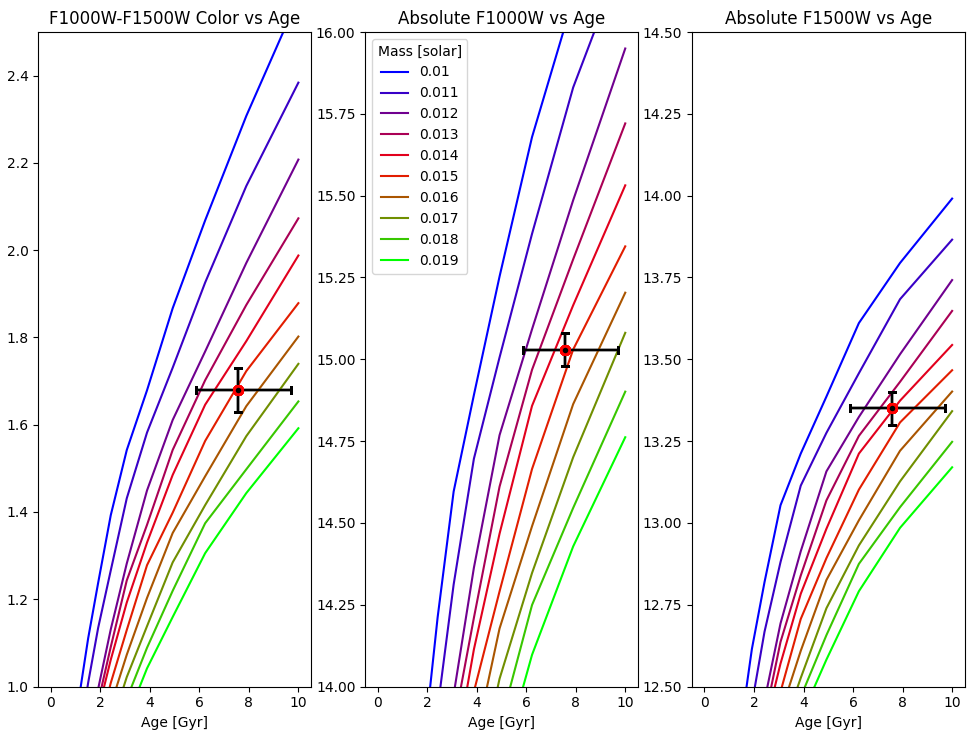}
    \caption{{\it Left:} Inferred mass for the candidate companion \meadb. Corner plot of the inferred age, mass and error scaling factor, log\_f, obtained from a MCMC run using the measured F1000W and F1500W magnitudes, the \mead\, distance, the prior age distribution obtained by modeling the white dwarf SED, as well as using the ATMO2020 (strong non-equilibrium) evolutionary models. {\it Right:} Plots of the F1000-F1500W, absolute F1000W and F1500W magnitudes versus age filled with lines of iso-masses predicted by the same ATMO2020 models overplotted with the MCMC inferred mass for \meadb.}
    \label{fig:mass}
\end{figure}

The brown dwarf inferred physical properties are mass $ = 0.014^{+0.002}_{-0.003}$\,\Msun (or 11.5 to 16.8 \Mjup), age $ = 7.6^{+1.7}_{-2.2}$\,Gyr and $T_{\rm eff} = 343^{+7}_{-11}$\,K. Note that the inferred age is about 1\,Gyr older, but within uncertainties, than the prior age of the system estimated from the white dwarf SED fit alone. This comes from the additional constraints provided by the brown dwarf photometry.
Figure\,\ref{fig:mass} shows the corner plot for the retrieved mass and age distributions as well as isomass model tracks in the magnitude vs. age or color vs. age diagrams. 

\section{False Positives Analysis} \label{sec:falsepositives}

Due to the high JWST sensitivity, even the short exposures used in this survey reveal a large fraction of galaxies compared to unresolved point sources. Also, most of the extended objects appear red (bright in F1500W compared to F1000W). With the somewhat coarser spatial resolution of MIRI compared to near infrared bands, these red galaxies, especially those at low SNR, could be confused with the unresolved point sources like the giant exoplanets or BDs that we seek.

To estimate the false positive rate, we therefore studied the occurrence rate of red unresolved sources in the entire MEAD survey. We adopted two criteria: the F1000W-F1500W color and the source FWHM. We retained objects that are both much redder than the stellar track and whose PSF is consistent with being unresolved. We defined as red any object whose measured color is F1000W$-$F1500$ \geq +0.80$\,mag. As for the FWHM criterion to select unresolved sources, its value depends on the flux SNR of each source. To determine that relation, we ran simulations of point-source PSFs with various amounts of readout and photon noises. These reproduced very well actual observations when overlaid in a FWHM versus SNR diagram. They also outline a FWHM upper boundary above which no data point was produced by the PSF simulation, thus was likely a resolved, extended object. For the F1000W band, this threshold varies continuously from FWHM=3.55 pixels at SNR$\geq100$ to FWHM=4.98 pixels at SNR$=25$. For the F1500W band, it goes from FWHM=5.15 to 8.67 pixels. Sources below those FWHM thresholds were classified as unresolved point sources and those with F1000W$-$F1500$ \geq +0.80$\,mag, as red.

With these FWHM and color criteria applied to our whole MEAD survey (56 fields), the density of red, unresolved point-sources could be established. On average, there were 12.4 red point-sources per MIRI field of view, down to the lowest flux SNR. That density is lower (10.6/FOV) when counting only red point-sources with SNR equal or larger than that of our only brown dwarf candidate. We did not see a correlation between the rate of faint red dots and the galactic latitude which suggest that many must be extra-galactic sources. Adopting 11376 sq. arcsec/FOV and an apparent separation of 2.0 arcsec for our candidate companion, we expect to find $1.2\times10^{-2}$ red point-sources within the same angular separation as our candidate, for any given observation, if these were randomly distributed. That figure is for the \mead\, field alone but the false-positive rate increases linearly with the number of fields surveyed. The expected false-positive occurrence reaches 0.66 when considering the whole 56-target MEAD survey. If we assume that the occurrence of false positives in a survey follows a Poisson distribution with an expected value of 0.66, then we should observe $\geq1$ false positive companions 48\% of times, in a given survey, or no false positive, 52\% of times. In other words, given that \meadb\, is the only believable companion candidate (there is no red, point-source candidate brighter and closer), the odds are almost equal that our candidate is a false positive or a true companion. Given the potentially unique nature of this candidate (one of the coldest white dwarf companions known to date), we believe that the 50/50 odds of it being real were sufficient to warrant a publication. However, we are cautious that \meadb\, remains only a candidate companion until further observations to confirm its nature are secured.


\section{Discussion} \label{sec:discussion}

The MEAD survey targeted nearby ($\leq 25$\,pc) white dwarfs using short MIRI imaging snapshots at 10\,$\mu$m and 15\,$\mu$m with the goal to detect planetary-mass companions, either through infrared excess emission or direct imaging. This JWST survey program executed 56 out of the 230 defined observations in the MEAD sample. Upon inspection of the images, a candidate companion around \mead\, was revealed at an apparent separation of $1.953\pm0.002$\arcsec (40.0 au). The shape of the candidate PSF is consistent with it being an unresolved point-source. If physically associated with the white dwarf, the color and both F1000W and F1500W absolute magnitudes fall along the sequence of ATMO2020 models (strong non-equilibrium) and other BDs observed with JWST. \meadb\, would be an old, $7.6^{+1.7}_{-2.2}$\,Gyr, brown dwarf with a mass of $ = 0.014^{+0.002}_{-0.003}$\,\Msun near the deuterium-burning boundary and a temperature of $343^{+7}_{-11}$\,K, surely in the Y-spectral type regime. The candidate brown dwarf \meadb, if confirmed, could become the coldest example of a population of known substellar objects gravitationally bound to a white dwarf. It would be only the second white-dwarf + Y-dwarf binary discovered after WD\,0806-661\,B \citep{luhman.2011,leggett.2017}.

The projected separation of 40\,au converted to when the primary star was on the main-sequence is roughly $\frac{ {\rm mass}_{\rm WD}}  {{\rm mass}_{\rm MS}} \times 40$\,au = $\frac{0.6}{1.35^{+0.4}_{-0.25}} \times 40$\,au $= 18\pm5$\,au. This is similar to the separation of Uranus in our own solar system. Also, about a dozen known directly imaged exoplanets have been found at similar apparent separations between 13 and 30\,au\footnote{NASA Exoplanet Archive, \url{https://exoplanetarchive.ipac.caltech.edu/}.}. Furthermore, a subgroup of five have masses close to that of \meadb\, between 10 and 20 \Mjup (i.e. HR~2562~b, HR~8799~d, HIP~81208~C~b, HIP~99770~b and HR~8799~e). But the known system matching most closely \meadb\, is HR~2562~b \citep{konopacky.2016}. It has an apparent separation of $27\pm3$\,au, an estimated mass of 10 to 14 \Mjup \citep{godoy.2024}, and its host spectral type of F5V corresponds to a host mass very similar to the main-sequence progenitor of \mead.


\subsection{Confirmation with Proper Motion}

Given the change in position caused by proper motion, a follow-up observation to establish common proper motion is absolutely required to confirm that the candidate companion \meadb\, is physically associated with \mead. The earliest feasible observation with MIRI imaging would be during JWST cycle 5, nearly three years after the discovery epoch of 2024 February 8th. Given the proper motion of \mead\, measured by Gaia (DR3) of $\Delta\alpha = 68.19\pm0.03$\,mas, $\Delta\delta = -39.61\pm0.03$\,mas, the expected motion ($\geq237$\,mas) would be larger than two MIRI pixel (110\,mas), thus easily measured.

\subsection{A rare ultracool atmosphere template}

If confirmed, the candidate companion \meadb\, would join a group of rare known ultracool BDs with $T_{\rm eff} \leq 350$~K including WISE\,J085510.83-071442.5 at 260\,K \citep{luhman.2014,leggett.2021}, WISE\,J033605.05-014350.4\,B at $T_{\rm eff}=325$\,K \citep{calissendorff.2023,leggett.2023} and WISEA\,J083011.95+283716.0 at $T_{\rm eff}=350$\,K \citep[K. Matuszewska submitted]{albert.2025}. For context, less than 50 BDs with spectral type Y (T$_{\rm eff}\leq500$\,K) are currently known, of which only 17 have T$_{\rm eff}\leq400$\,K \citep{kirkpatrick.2021}. Objects that cold are key to understanding atmospheres in the temperature regime where water clouds start forming. Atmosphere models including water clouds such as those of \cite{lacy.2023} still have difficulties reproducing WISE\,J085510.83-071442.5 and need a larger population of such cold BDs to establish and reproduce statistical trends. The recent JWST detection of the massive exoplanet WD\,1856+534\,b \citep[mass $= 5.2^{+0.7}_{-0.8}$\,\Mjup, $T_{\rm eff}=185\pm5$\,K]{limbach.2025} shows that looking for companions around white dwarfs in the mid-IR is a promising way to identify such a population of ultracool objects.

\subsection{Brown dwarf binary occurrence rate}

Radial velocity surveys have shown a deficit of brown dwarfs orbiting main-sequence stars at separations of $\leq3$\,AU, the so-called brown dwarf desert \citep{marcy.2000, grether.2006}. However, at larger separations, similar to where our candidate \meadb\, is found, BDs around main-sequence stars do occur, at a rate of roughly 1-3\% \citep{gizis.2001,mccarthy.2004,metchev.2009,vigan.2017}. As for the number of brown dwarf companions to white dwarf primaries, it is relatively small. A recent search by \cite{bravo.2025} (and references therein) has brought the list of resolved, widely separated systems ($\geq50$\,au) to more than 50. Also, less than a dozen BD+WD with short orbital periods ($P\leq2$\,h) are known, indicative of being post-common envelope binaries (e.g. \cite{rappaport.2017, longstaff.2019}) thus likely have different formation pathways compared to widely separated BDs. However, statistical studies of photometrically-selected samples of WDs coupled to infrared surveys (WISE or 2MASS) to look for infrared excess emission suggest that the BD+WD binary occurrence rate ranges between 0.5 and 2\% \citep{farihi.2005, steele.2011, girven.2011}, so similar to what is found around main-sequence stars.

Preliminary analysis of our MEAD survey points to the discovery of a single candidate BD in a survey of 56 white dwarfs. Using a binomial distribution as in \cite{burgasser.2003}, this suggests a BD occurrence rate around white dwarfs of $f_{\rm BD} = 1.8^{+3.9}_{-0.5}$\,\%, similar to that found around main-sequence stars. These numbers may change as analysis of the full MEAD sample of white dwarfs is still underway and may yield additional candidates which would increase the occurrence rate. Also, a quantitative analysis of the achieved contrasts is needed to determine the mass sensitivity limit which we roughly estimate at 4-8\,\Mjup. If confirmed, that sensitivity would mean that the MEAD survey reaches contrasts making it complete in the BD mass regime ($13$\,\Mjup\,$\leq$\,mass\,$\leq 75$\,\Mjup).










\subsection{Companion formation scenario}

The companion occurrence rate around main-sequence stars reaches a minimum in the BD regime, roughly between 10 and 50 \Mjup, supported both from radial-velocity studies \citep{grether.2006} and direct imaging surveys \citep{reggiani.2016}. This led \cite{reggiani.2016} to suggest that two formation processes are at play when producing companions: protoplanetary disk fragmentation producing planets, protostellar core fragmentation producing sub-stellar objects. Also, upon measuring a shift to higher masses for this occurrence minimum around more massive stars, \cite{duchene.2023} suggest that the companion to primary mass ratio, q, rather than the absolute companion mass, may be the deciding parameter determining the formation pathway, with q$\leq0.02$ being the threshold under which disk fragmentation formation is at play. Using the estimated main-sequence progenitor mass for \mead\,of $1.35^{+0.40}_{-0.25}$\,\Msun (Fig. \ref{fig:wdproperties}) and the inferred mass for \meadb\, of $0.014^{+0.002}_{-0.003}$\,\Msun, the companion-to-primary mass ratio is q=$0.010\pm0.003$. Both this low mass ratio being under the q$\leq0.02$ threshold and the companion mass being near the lower bound of the BD desert favor \meadb\,having formed as the result of protoplanetary disk fragmentation rather than protostellar core fragmentation.

\section{Conclusions} \label{sec:conclusion}

A quick analysis of the MEAD survey targeting giant planets around white dwarfs has yielded its first discovery: the directly imaged candidate brown dwarf companion, \meadb. F1000W and F1500W MIRI photometry puts this candidate right along the ATMO2020 brown dwarf model track in the color-magnitude diagrams. Armed with a prior age estimate obtained from modelling the white dwarf SED model suggest that \meadb\, is a rare ultracool world with T$_{\rm eff} = 343^{+7}_{-11}$\,K, about $7.6^{+1.7}_{-2.2}$\,Gyr old with a mass of $ = 0.014^{+0.002}_{-0.003}$\,\Msun\, ($\sim14$\Mjup). 

We emphasize that \meadb\, remains a candidate companion, for now. A second epoch observation is needed to measure the companion proper motion and confirm whether \meadb\ is indeed physically associated with the white dwarf. \meadb\, the ATMO2020 models predict an apparent brightness of $J_{\rm Vega} = 23.5-27.0$, equivalent H-band magnitude and at least 3 magnitudes fainter in K-band. A follow-up from the ground could be attempted in imaging on an 8-meter class telescope in the southern hemisphere but would not guarantee a detection. However, galaxies are expected to have $J_{\rm Vega} \approx 20$ so a clear detection could disprove \meadb\, as a false-positive. Characterizing the full spectral energy distribution of the brown dwarf would only be possible with JWST. Our false positive analysis of background unresolved red sources in the 56-field MEAD survey indicates that the odds are equal that this is a real ultracool companion or an unresolved false-positive red dot. Based on an empirical PSF fitting, we could at least confirm that \meadb\, is consistent with being an unresolved point source, rather than a resolved background galaxy. 

A careful analysis of all images in the MEAD survey is currently underway to search for resolved and unresolved giant planetary companions. In particular, establishing a list of reliable direct imaging candidates will require a careful empirical PSF fitting analysis of all currently unresolved white dwarfs in the survey to look for high-contrast binaries. In parallel, securing infrared excess emission candidates will call for accurate modeling of the white dwarfs SED. The MEAD survey is expected to be sensitive to planetary-mass objects down to 4-8\Mjup, depending whether these are resolved or detected as excess emission.


\begin{acknowledgments}

This work is based on observations made with the NASA/ESA/CSA James Webb Space Telescope. The data were obtained from the Mikulski Archive for Space Telescopes at the Space Telescope Science Institute, which is operated by the Association of Universities for Research in Astronomy, Inc., under NASA contract NAS 5-03127 for JWST. These observations are associated with program \#3964. Support for this program was provided through a grant from the STScI under NASA contract NAS 5-03127. L.A. warmly thanks the Canadian Space Agency for its financial support through grant 23JWGO2B07. M.K. acknowledges support by the NSF under grant AST-2205736 and NASA under grant Nos. 80NSSC22K0479, 80NSSC24K0380, and 80NSSC24K0436. M.D.F. is supported by an NSF Astronomy and Astrophysics Postdoctoral Fellowship under award AST-2303911.

\end{acknowledgments}


\bibliography{manuscript}{}

\begin{thebibliography}{}
\expandafter\ifx\csname natexlab\endcsname\relax\def\natexlab#1{#1}\fi
\providecommand{\url}[1]{\href{#1}{#1}}
\providecommand{\dodoi}[1]{doi:~\href{http://doi.org/#1}{\nolinkurl{#1}}}
\providecommand{\doeprint}[1]{\href{http://ascl.net/#1}{\nolinkurl{http://ascl.net/#1}}}
\providecommand{\doarXiv}[1]{\href{https://arxiv.org/abs/#1}{\nolinkurl{https://arxiv.org/abs/#1}}}

\bibitem[{L. {Albert} {et~al.}(2025){Albert}, {Leggett}, {Calissendorff},
  {Vandal}, {Kirkpatrick}, {Bardalez Gagliuffi}, {De Furio}, {Meyer},
  {Beichman}, {Burgasser}, {Cushing}, {Faherty}, {Fontanive}, {Gelino},
  {Gizis}, {Greenbaum}, {Martinache}, {N'Diaye}, {Pope}, {Roellig}, {Sahlmann},
  {Sivaramakrishnan}, \& {Ygouf}}]{albert.2025}
{Albert}, L., {Leggett}, S.~K., {Calissendorff}, P., {et~al.} 2025,
  \bibinfo{title}{{JWST 1.5 {\ensuremath{\mu}}m and 4.8 {\ensuremath{\mu}}m
  Photometry of Y Dwarfs},} \aj, 169, 163, \dodoi{10.3847/1538-3881/adadf9}

\bibitem[{C.~A.~L. {Bailer-Jones} {et~al.}(2021){Bailer-Jones}, {Rybizki},
  {Fouesneau}, {Demleitner}, \& {Andrae}}]{Bailer21}
{Bailer-Jones}, C.~A.~L., {Rybizki}, J., {Fouesneau}, M., {Demleitner}, M., \&
  {Andrae}, R. 2021, \bibinfo{title}{{Estimating Distances from Parallaxes. V.
  Geometric and Photogeometric Distances to 1.47 Billion Stars in Gaia Early
  Data Release 3},} \aj, 161, 147, \dodoi{10.3847/1538-3881/abd806}

\bibitem[{S.~D. {Barber} {et~al.}(2012){Barber}, {Patterson}, {Kilic},
  {Leggett}, {Dufour}, {Bloom}, \& {Starr}}]{barber2012}
{Barber}, S.~D., {Patterson}, A.~J., {Kilic}, M., {et~al.} 2012,
  \bibinfo{title}{{The Frequency of Debris Disks at White Dwarfs},} \apj, 760,
  26, \dodoi{10.1088/0004-637X/760/1/26}

\bibitem[{E.~E. {Becklin} {et~al.}(2005){Becklin}, {Farihi}, {Jura}, {Song},
  {Weinberger}, \& {Zuckerman}}]{becklin.2005}
{Becklin}, E.~E., {Farihi}, J., {Jura}, M., {et~al.} 2005, \bibinfo{title}{{A
  Dusty Disk around GD 362, a White Dwarf with a Uniquely High Photospheric
  Metal Abundance},} \apjl, 632, L119, \dodoi{10.1086/497826}

\bibitem[{E.~E. {Becklin} \& B. {Zuckerman}(1988){Becklin} \&
  {Zuckerman}}]{becklin.1988}
{Becklin}, E.~E., \& {Zuckerman}, B. 1988, \bibinfo{title}{{A low-temperature
  companion to a white dwarf star},} \nat, 336, 656, \dodoi{10.1038/336656a0}

\bibitem[{A. {B{\'e}dard} {et~al.}(2020){B{\'e}dard}, {Bergeron}, {Brassard},
  \& {Fontaine}}]{Bedard20}
{B{\'e}dard}, A., {Bergeron}, P., {Brassard}, P., \& {Fontaine}, G. 2020,
  \bibinfo{title}{{On the Spectral Evolution of Hot White Dwarf Stars. I. A
  Detailed Model Atmosphere Analysis of Hot White Dwarfs from SDSS DR12},}
  \apj, 901, 93, \dodoi{10.3847/1538-4357/abafbe}

\bibitem[{S.~A. {Beiler} {et~al.}(2024){Beiler}, {Cushing}, {Kirkpatrick},
  {Schneider}, {Mukherjee}, {Marley}, {Marocco}, \& {Smart}}]{beiler.2024}
{Beiler}, S.~A., {Cushing}, M.~C., {Kirkpatrick}, J.~D., {et~al.} 2024,
  \bibinfo{title}{{Precise Bolometric Luminosities and Effective Temperatures
  of 23 late-T and Y dwarfs Obtained with JWST},} arXiv e-prints,
  arXiv:2407.08518, \dodoi{10.48550/arXiv.2407.08518}

\bibitem[{P. {Bergeron} {et~al.}(2019){Bergeron}, {Dufour}, {Fontaine},
  {Coutu}, {Blouin}, {Genest-Beaulieu}, {B{\'e}dard}, \&
  {Rolland}}]{Bergeron2019}
{Bergeron}, P., {Dufour}, P., {Fontaine}, G., {et~al.} 2019,
  \bibinfo{title}{{On the Measurement of Fundamental Parameters of White Dwarfs
  in the Gaia Era},} \apj, 876, 67, \dodoi{10.3847/1538-4357/ab153a}

\bibitem[{E. {Bertin} \& S. {Arnouts}(1996){Bertin} \& {Arnouts}}]{bertin.1996}
{Bertin}, E., \& {Arnouts}, S. 1996, \bibinfo{title}{{SExtractor: Software for
  source extraction.},} \aaps, 117, 393, \dodoi{10.1051/aas:1996164}

\bibitem[{S. {Blouin} {et~al.}(2024){Blouin}, {Kilic}, {Albert},
  {Azartash-Namin}, \& {Dufour}}]{blouin.2024}
{Blouin}, S., {Kilic}, M., {Albert}, L., {Azartash-Namin}, B., \& {Dufour}, P.
  2024, \bibinfo{title}{{JWST Resolves Collision-induced Absorption Features in
  White Dwarfs},} \apj, 976, 218, \dodoi{10.3847/1538-4357/ad863b}

\bibitem[{S. {Blouin} {et~al.}(2017){Blouin}, {Kowalski}, \&
  {Dufour}}]{Blouin2017}
{Blouin}, S., {Kowalski}, P.~M., \& {Dufour}, P. 2017,
  \bibinfo{title}{{Pressure Distortion of the H$_{2}$-He Collision-induced
  Absorption at the Photosphere of Cool White Dwarf Stars},} \apj, 848, 36,
  \dodoi{10.3847/1538-4357/aa8ad6}

\bibitem[{A. {Bravo} {et~al.}(2025){Bravo}, {Schneider}, {Casewell},
  {Rothermich}, {Faherty}, {French}, {Bickle}, {Meisner}, {Kirkpatrick},
  {Kuchner}, {Burgasser}, {Marocco}, {Debes}, {Sainio}, {Gramaize}, {Kiwy},
  {Ja{\l}owiczor}, \& {Abdullahi}}]{bravo.2025}
{Bravo}, A., {Schneider}, A.~C., {Casewell}, S., {et~al.} 2025,
  \bibinfo{title}{{New Ultracool Companions to Nearby White Dwarfs},} \aj, 169,
  100, \dodoi{10.3847/1538-3881/ad9b9b}

\bibitem[{A.~J. {Burgasser} {et~al.}(2003){Burgasser}, {Kirkpatrick}, {Reid},
  {Brown}, {Miskey}, \& {Gizis}}]{burgasser.2003}
{Burgasser}, A.~J., {Kirkpatrick}, J.~D., {Reid}, I.~N., {et~al.} 2003,
  \bibinfo{title}{{Binarity in Brown Dwarfs: T Dwarf Binaries Discovered with
  the Hubble Space Telescope Wide Field Planetary Camera 2},} \apj, 586, 512,
  \dodoi{10.1086/346263}

\bibitem[{A. {Burrows} {et~al.}(1997){Burrows}, {Marley}, {Hubbard}, {Lunine},
  {Guillot}, {Saumon}, {Freedman}, {Sudarsky}, \& {Sharp}}]{burrows.1997}
{Burrows}, A., {Marley}, M., {Hubbard}, W.~B., {et~al.} 1997,
  \bibinfo{title}{{A Nongray Theory of Extrasolar Giant Planets and Brown
  Dwarfs},} \apj, 491, 856, \dodoi{10.1086/305002}

\bibitem[{P. {Calissendorff} {et~al.}(2023){Calissendorff}, {De Furio},
  {Meyer}, {Albert}, {Aganze}, {Ali-Dib}, {Bardalez Gagliuffi}, {Baron},
  {Beichman}, {Burgasser}, {Cushing}, {Faherty}, {Fontanive}, {Gelino},
  {Gizis}, {Greenbaum}, {Kirkpatrick}, {Leggett}, {Martinache}, {Mary},
  {N'Diaye}, {Pope}, {Roellig}, {Sahlmann}, {Sivaramakrishnan}, {Thorngren},
  {Ygouf}, \& {Vandal}}]{calissendorff.2023}
{Calissendorff}, P., {De Furio}, M., {Meyer}, M., {et~al.} 2023,
  \bibinfo{title}{{JWST/NIRCam Discovery of the First Y+Y Brown Dwarf Binary:
  WISE J033605.05-014350.4},} \apjl, 947, L30, \dodoi{10.3847/2041-8213/acc86d}

\bibitem[{S.~L. {Casewell} {et~al.}(2018){Casewell}, {Braker}, {Parsons},
  {Hermes}, {Burleigh}, {Belardi}, {Chaushev}, {Finch}, {Roy}, {Littlefair},
  {Goad}, \& {Dennihy}}]{Casewell18}
{Casewell}, S.~L., {Braker}, I.~P., {Parsons}, S.~G., {et~al.} 2018,
  \bibinfo{title}{{The first sub-70 min non-interacting WD-BD system:
  EPIC212235321},} \mnras, 476, 1405, \dodoi{10.1093/mnras/sty245}

\bibitem[{J.~D. {Cummings} {et~al.}(2018){Cummings}, {Kalirai}, {Tremblay},
  {Ramirez-Ruiz}, \& {Choi}}]{cummings.2018}
{Cummings}, J.~D., {Kalirai}, J.~S., {Tremblay}, P.~E., {Ramirez-Ruiz}, E., \&
  {Choi}, J. 2018, \bibinfo{title}{{The White Dwarf Initial-Final Mass Relation
  for Progenitor Stars from 0.85 to 7.5 M $_{{\ensuremath{\odot}}}$},} \apj,
  866, 21, \dodoi{10.3847/1538-4357/aadfd6}

\bibitem[{M. {De Furio} {et~al.}(2023){De Furio}, {Lew}, {Beichman}, {Roellig},
  {Bryden}, {Ciardi}, {Meyer}, {Rieke}, {Greenbaum}, {Leisenring},
  {Llop-Sayson}, {Ygouf}, {Albert}, {Boyer}, {Eisenstein}, {Hodapp}, {Horner},
  {Johnstone}, {Kelly}, {Misselt}, {Rieke}, {Stansberry}, \&
  {Young}}]{defurio.2023}
{De Furio}, M., {Lew}, B., {Beichman}, C., {et~al.} 2023, \bibinfo{title}{{JWST
  Observations of the Enigmatic Y-Dwarf WISE 1828+2650. I. Limits to a Binary
  Companion},} \apj, 948, 92, \dodoi{10.3847/1538-4357/acbf1e}

\bibitem[{J.~H. {Debes} {et~al.}(2011){Debes}, {Hoard}, {Wachter}, {Leisawitz},
  \& {Cohen}}]{debes11}
{Debes}, J.~H., {Hoard}, D.~W., {Wachter}, S., {Leisawitz}, D.~T., \& {Cohen},
  M. 2011, \bibinfo{title}{{The WIRED Survey. II. Infrared Excesses in the SDSS
  DR7 White Dwarf Catalog},} \apjs, 197, 38, \dodoi{10.1088/0067-0049/197/2/38}

\bibitem[{G. {Duch{\^e}ne} {et~al.}(2023){Duch{\^e}ne}, {Oon}, {De Rosa},
  {Kantorski}, {Coy}, {Wang}, {Thomas}, {Patience}, {Pueyo}, {Nielsen}, \&
  {Konopacky}}]{duchene.2023}
{Duch{\^e}ne}, G., {Oon}, J.~T., {De Rosa}, R.~J., {et~al.} 2023,
  \bibinfo{title}{{A low-mass companion desert among intermediate-mass visual
  binaries: The scaled-up counterpart to the brown dwarf desert},} \mnras, 519,
  778, \dodoi{10.1093/mnras/stac3527}

\bibitem[{T.~J. {Dupuy} \& M.~C. {Liu}(2017){Dupuy} \& {Liu}}]{dupuy.2017}
{Dupuy}, T.~J., \& {Liu}, M.~C. 2017, \bibinfo{title}{{Individual Dynamical
  Masses of Ultracool Dwarfs},} \apjs, 231, 15,
  \dodoi{10.3847/1538-4365/aa5e4c}

\bibitem[{J. {Farihi}(2016){Farihi}}]{farihi16}
{Farihi}, J. 2016, \bibinfo{title}{{Circumstellar debris and pollution at white
  dwarf stars},} \nar, 71, 9, \dodoi{10.1016/j.newar.2016.03.001}

\bibitem[{J. {Farihi} {et~al.}(2005){Farihi}, {Becklin}, \&
  {Zuckerman}}]{farihi.2005}
{Farihi}, J., {Becklin}, E.~E., \& {Zuckerman}, B. 2005,
  \bibinfo{title}{{Low-Luminosity Companions to White Dwarfs},} \apjs, 161,
  394, \dodoi{10.1086/444362}

\bibitem[{J.~R. {French} {et~al.}(2023){French}, {Casewell}, {Dupuy}, {Debes},
  {Manjavacas}, {Martin}, \& {Xu}}]{French23}
{French}, J.~R., {Casewell}, S.~L., {Dupuy}, T.~J., {et~al.} 2023,
  \bibinfo{title}{{Discovery of a resolved white dwarf-brown dwarf binary with
  a small projected separation: SDSS J222551.65+001637.7AB},} \mnras, 519,
  5008, \dodoi{10.1093/mnras/stac3807}

\bibitem[{J.~R. {French} {et~al.}(2024){French}, {Casewell}, {Amaro},
  {Lothringer}, {Mayorga}, {Littlefair}, {Lew}, {Zhou}, {Apai}, {Marley},
  {Parmentier}, \& {Tan}}]{French24}
{French}, J.~R., {Casewell}, S.~L., {Amaro}, R.~C., {et~al.} 2024,
  \bibinfo{title}{{The only inflated brown dwarf in an eclipsing white
  dwarf-brown dwarf binary: WD1032+011B},} \mnras, 534, 2244,
  \dodoi{10.1093/mnras/stae2121}

\bibitem[{J. {Girven} {et~al.}(2011){Girven}, {G{\"a}nsicke}, {Steeghs}, \&
  {Koester}}]{girven.2011}
{Girven}, J., {G{\"a}nsicke}, B.~T., {Steeghs}, D., \& {Koester}, D. 2011,
  \bibinfo{title}{{DA white dwarfs in Sloan Digital Sky Survey Data Release 7
  and a search for infrared excess emission},} \mnras, 417, 1210,
  \dodoi{10.1111/j.1365-2966.2011.19337.x}

\bibitem[{J.~E. {Gizis} {et~al.}(2001){Gizis}, {Kirkpatrick}, {Burgasser},
  {Reid}, {Monet}, {Liebert}, \& {Wilson}}]{gizis.2001}
{Gizis}, J.~E., {Kirkpatrick}, J.~D., {Burgasser}, A., {et~al.} 2001,
  \bibinfo{title}{{Substellar Companions to Main-Sequence Stars: No Brown Dwarf
  Desert at Wide Separations},} \apjl, 551, L163, \dodoi{10.1086/320017}

\bibitem[{N. {Godoy} {et~al.}(2024){Godoy}, {Choquet}, {Serabyn}, {Danielski},
  {Stolker}, {Charnay}, {Hinkley}, {Lagage}, {Ressler}, {Tremblin}, \&
  {Vigan}}]{godoy.2024}
{Godoy}, N., {Choquet}, E., {Serabyn}, E., {et~al.} 2024, \bibinfo{title}{{A
  new atmospheric characterization of the sub-stellar companion HR 2562 B with
  JWST/MIRI observations},} \aap, 689, A185,
  \dodoi{10.1051/0004-6361/202449951}

\bibitem[{D. {Grether} \& C.~H. {Lineweaver}(2006){Grether} \&
  {Lineweaver}}]{grether.2006}
{Grether}, D., \& {Lineweaver}, C.~H. 2006, \bibinfo{title}{{How Dry is the
  Brown Dwarf Desert? Quantifying the Relative Number of Planets, Brown Dwarfs,
  and Stellar Companions around Nearby Sun-like Stars},} \apj, 640, 1051,
  \dodoi{10.1086/500161}

\bibitem[{M.~A. {Hollands} {et~al.}(2018){Hollands}, {Tremblay},
  {G{\"a}nsicke}, {Gentile-Fusillo}, \& {Toonen}}]{Hollands2018}
{Hollands}, M.~A., {Tremblay}, P.~E., {G{\"a}nsicke}, B.~T., {Gentile-Fusillo},
  N.~P., \& {Toonen}, S. 2018, \bibinfo{title}{{The Gaia 20 pc white dwarf
  sample},} \mnras, 480, 3942, \dodoi{10.1093/mnras/sty2057}

\bibitem[{M. {Kilic} {et~al.}(2005){Kilic}, {von Hippel}, {Leggett}, \&
  {Winget}}]{kilic.2005}
{Kilic}, M., {von Hippel}, T., {Leggett}, S.~K., \& {Winget}, D.~E. 2005,
  \bibinfo{title}{{Excess Infrared Radiation from the Massive DAZ White Dwarf
  GD 362: A Debris Disk?},} \apjl, 632, L115, \dodoi{10.1086/497825}

\bibitem[{R. {Kiman} {et~al.}(2022){Kiman}, {Xu}, {Faherty}, {Gagn{\'e}},
  {Angus}, {Brandt}, {Casewell}, \& {Cruz}}]{kiman.2022}
{Kiman}, R., {Xu}, S., {Faherty}, J.~K., {et~al.} 2022,
  \bibinfo{title}{{wdwarfdate: A Python Package to Derive Bayesian Ages of
  White Dwarfs},} \aj, 164, 62, \dodoi{10.3847/1538-3881/ac7788}

\bibitem[{J.~D. {Kirkpatrick} {et~al.}(2021){Kirkpatrick}, {Gelino}, {Faherty},
  {Meisner}, {Caselden}, {Schneider}, {Marocco}, {Cayago}, {Smart},
  {Eisenhardt}, {Kuchner}, {Wright}, {Cushing}, {Allers}, {Bardalez Gagliuffi},
  {Burgasser}, {Gagn{\'e}}, {Logsdon}, {Martin}, {Ingalls}, {Lowrance},
  {Abrahams}, {Aganze}, {Gerasimov}, {Gonzales}, {Hsu}, {Kamraj}, {Kiman},
  {Rees}, {Theissen}, {Ammar}, {Andersen}, {Beaulieu}, {Colin}, {Elachi},
  {Goodman}, {Gramaize}, {Hamlet}, {Hong}, {Jonkeren}, {Khalil}, {Martin},
  {Pendrill}, {Pumphrey}, {Rothermich}, {Sainio}, {Stenner}, {Tanner},
  {Th{\'e}venot}, {Voloshin}, {Walla}, {W{\k{e}}dracki}, \& {Backyard Worlds:
  Planet 9 Collaboration}}]{kirkpatrick.2021}
{Kirkpatrick}, J.~D., {Gelino}, C.~R., {Faherty}, J.~K., {et~al.} 2021,
  \bibinfo{title}{{The Field Substellar Mass Function Based on the Full-sky 20
  pc Census of 525 L, T, and Y Dwarfs},} \apjs, 253, 7,
  \dodoi{10.3847/1538-4365/abd107}

\bibitem[{Q.~M. {Konopacky} {et~al.}(2016){Konopacky}, {Rameau}, {Duch{\^e}ne},
  {Filippazzo}, {Giorla Godfrey}, {Marois}, {Nielsen}, {Pueyo}, {Rafikov},
  {Rice}, {Wang}, {Ammons}, {Bailey}, {Barman}, {Bulger}, {Bruzzone},
  {Chilcote}, {Cotten}, {Dawson}, {De Rosa}, {Doyon}, {Esposito}, {Fitzgerald},
  {Follette}, {Goodsell}, {Graham}, {Greenbaum}, {Hibon}, {Hung}, {Ingraham},
  {Kalas}, {Lafreni{\`e}re}, {Larkin}, {Macintosh}, {Maire}, {Marchis},
  {Marley}, {Matthews}, {Metchev}, {Millar-Blanchaer}, {Oppenheimer}, {Palmer},
  {Patience}, {Perrin}, {Poyneer}, {Rajan}, {Rantakyr{\"o}}, {Savransky},
  {Schneider}, {Sivaramakrishnan}, {Song}, {Soummer}, {Thomas}, {Wallace},
  {Ward-Duong}, {Wiktorowicz}, \& {Wolff}}]{konopacky.2016}
{Konopacky}, Q.~M., {Rameau}, J., {Duch{\^e}ne}, G., {et~al.} 2016,
  \bibinfo{title}{{Discovery of a Substellar Companion to the Nearby Debris
  Disk Host HR 2562},} \apjl, 829, L4, \dodoi{10.3847/2041-8205/829/1/L4}

\bibitem[{H. {K{\"u}hnle} {et~al.}(2025){K{\"u}hnle}, {Patapis},
  {Molli{\`e}re}, {Tremblin}, {Matthews}, {Glauser}, {Whiteford}, {Vasist},
  {Absil}, {Barrado}, {Min}, {Lagage}, {Waters}, {Guedel}, {Henning},
  {Vandenbussche}, {Baudoz}, {Decin}, {Pye}, {Royer}, {van Dishoeck},
  {{\"O}stlin}, {Ray}, \& {Wright}}]{kuhnle.2025}
{K{\"u}hnle}, H., {Patapis}, P., {Molli{\`e}re}, P., {et~al.} 2025,
  \bibinfo{title}{{Water depletion and $^{15}$NH$_{3}$ in the atmosphere of the
  coldest brown dwarf observed with JWST/MIRI},} \aap, 695, A224,
  \dodoi{10.1051/0004-6361/202452547}

\bibitem[{B. {Lacy} \& A. {Burrows}(2023){Lacy} \& {Burrows}}]{lacy.2023}
{Lacy}, B., \& {Burrows}, A. 2023, \bibinfo{title}{{Self-consistent Models of Y
  Dwarf Atmospheres with Water Clouds and Disequilibrium Chemistry},} \apj,
  950, 8, \dodoi{10.3847/1538-4357/acc8cb}

\bibitem[{S.~K. {Leggett} \& P. {Tremblin}(2023){Leggett} \&
  {Tremblin}}]{leggett.2023}
{Leggett}, S.~K., \& {Tremblin}, P. 2023, \bibinfo{title}{{The First Y Dwarf
  Data from JWST Show that Dynamic and Diabatic Processes Regulate Cold Brown
  Dwarf Atmospheres},} \apj, 959, 86, \dodoi{10.3847/1538-4357/acfdad}

\bibitem[{S.~K. {Leggett} {et~al.}(2017){Leggett}, {Tremblin}, {Esplin},
  {Luhman}, \& {Morley}}]{leggett.2017}
{Leggett}, S.~K., {Tremblin}, P., {Esplin}, T.~L., {Luhman}, K.~L., \&
  {Morley}, C.~V. 2017, \bibinfo{title}{{The Y-type Brown Dwarfs: Estimates of
  Mass and Age from New Astrometry, Homogenized Photometry, and Near-infrared
  Spectroscopy},} \apj, 842, 118, \dodoi{10.3847/1538-4357/aa6fb5}

\bibitem[{S.~K. {Leggett} {et~al.}(2021){Leggett}, {Tremblin}, {Phillips},
  {Dupuy}, {Marley}, {Morley}, {Schneider}, {Caselden}, {Guillaume}, \&
  {Logsdon}}]{leggett.2021}
{Leggett}, S.~K., {Tremblin}, P., {Phillips}, M.~W., {et~al.} 2021,
  \bibinfo{title}{{Measuring and Replicating the 1-20 {\ensuremath{\mu}}m
  Energy Distributions of the Coldest Brown Dwarfs: Rotating, Turbulent, and
  Nonadiabatic Atmospheres},} \apj, 918, 11, \dodoi{10.3847/1538-4357/ac0cfe}

\bibitem[{M. {Libralato} {et~al.}(2024){Libralato}, {Argyriou}, {Dicken},
  {Garc{\'\i}a Mar{\'\i}n}, {Guillard}, {Hines}, {Kavanagh}, {Kendrew}, {Law},
  {Noriega-Crespo}, \& {{\'A}lvarez-M{\'a}rquez}}]{libralato.2024}
{Libralato}, M., {Argyriou}, I., {Dicken}, D., {et~al.} 2024,
  \bibinfo{title}{{High-precision Astrometry and Photometry with the JWST/MIRI
  Imager},} \pasp, 136, 034502, \dodoi{10.1088/1538-3873/ad2551}

\bibitem[{M.~A. {Limbach} {et~al.}(2025){Limbach}, {Vanderburg}, {MacDonald},
  {Stevenson}, {Jenkins}, {Blouin}, {Rauscher}, {Bowens-Rubin}, {Gallo},
  {Mang}, {Morley}, {Sing}, {O'Connor}, {Venner}, \& {Xu}}]{limbach.2025}
{Limbach}, M.~A., {Vanderburg}, A., {MacDonald}, R.~J., {et~al.} 2025,
  \bibinfo{title}{{Thermal Emission and Confirmation of the Frigid White Dwarf
  Exoplanet WD 1856+534 b},} \apjl, 984, L28, \dodoi{10.3847/2041-8213/adc9ad}

\bibitem[{E.~F. {Linder} {et~al.}(2019){Linder}, {Mordasini}, {Molli{\`e}re},
  {Marleau}, {Malik}, {Quanz}, \& {Meyer}}]{linder.2019}
{Linder}, E.~F., {Mordasini}, C., {Molli{\`e}re}, P., {et~al.} 2019,
  \bibinfo{title}{{Evolutionary models of cold and low-mass planets: cooling
  curves, magnitudes, and detectability},} \aap, 623, A85,
  \dodoi{10.1051/0004-6361/201833873}

\bibitem[{E.~S. {Longstaff} {et~al.}(2019){Longstaff}, {Casewell}, {Wynn},
  {Page}, {Williams}, {Braker}, \& {Maxted}}]{longstaff.2019}
{Longstaff}, E.~S., {Casewell}, S.~L., {Wynn}, G.~A., {et~al.} 2019,
  \bibinfo{title}{{Signs of accretion in the white dwarf + brown dwarf binary
  NLTT5306},} \mnras, 484, 2566, \dodoi{10.1093/mnras/stz127}

\bibitem[{K.~L. {Luhman}(2014){Luhman}}]{luhman.2014}
{Luhman}, K.~L. 2014, \bibinfo{title}{{Discovery of a
  \raisebox{-0.5ex}\textasciitilde250 K Brown Dwarf at 2 pc from the Sun},}
  \apjl, 786, L18, \dodoi{10.1088/2041-8205/786/2/L18}

\bibitem[{K.~L. {Luhman} {et~al.}(2011){Luhman}, {Burgasser}, \&
  {Bochanski}}]{luhman.2011}
{Luhman}, K.~L., {Burgasser}, A.~J., \& {Bochanski}, J.~J. 2011,
  \bibinfo{title}{{Discovery of a Candidate for the Coolest Known Brown
  Dwarf},} \apjl, 730, L9, \dodoi{10.1088/2041-8205/730/1/L9}

\bibitem[{G.~W. {Marcy} \& R.~P. {Butler}(2000){Marcy} \&
  {Butler}}]{marcy.2000}
{Marcy}, G.~W., \& {Butler}, R.~P. 2000, \bibinfo{title}{{Planets Orbiting
  Other Suns},} \pasp, 112, 137, \dodoi{10.1086/316516}

\bibitem[{M.~S. {Marley} {et~al.}(2021){Marley}, {Saumon}, {Visscher}, {Lupu},
  {Freedman}, {Morley}, {Fortney}, {Seay}, {Smith}, {Teal}, \&
  {Wang}}]{marley.2021}
{Marley}, M.~S., {Saumon}, D., {Visscher}, C., {et~al.} 2021,
  \bibinfo{title}{{The Sonora Brown Dwarf Atmosphere and Evolution Models. I.
  Model Description and Application to Cloudless Atmospheres in Rainout
  Chemical Equilibrium},} \apj, 920, 85, \dodoi{10.3847/1538-4357/ac141d}

\bibitem[{E.~C. {Matthews} {et~al.}(2024){Matthews}, {Carter}, {Pathak},
  {Morley}, {Phillips}, {P.~M.}, {Feng}, {Bonse}, {Boogaard}, {Burt},
  {Crossfield}, {Douglas}, {Henning}, {Hom}, {Ko}, {Kasper}, {Lagrange}, {Petit
  dit de la Roche}, \& {Philipot}}]{matthews.2024}
{Matthews}, E.~C., {Carter}, A.~L., {Pathak}, P., {et~al.} 2024,
  \bibinfo{title}{{A temperate super-Jupiter imaged with JWST in the
  mid-infrared},} \nat, 633, 789, \dodoi{10.1038/s41586-024-07837-8}

\bibitem[{C. {McCarthy} \& B. {Zuckerman}(2004){McCarthy} \&
  {Zuckerman}}]{mccarthy.2004}
{McCarthy}, C., \& {Zuckerman}, B. 2004, \bibinfo{title}{{The Brown Dwarf
  Desert at 75-1200 AU},} \aj, 127, 2871, \dodoi{10.1086/383559}

\bibitem[{J. {McCleery} {et~al.}(2020){McCleery}, {Tremblay}, {Gentile
  Fusillo}, {Hollands}, {G{\"a}nsicke}, {Izquierdo}, {Toonen}, {Cunningham}, \&
  {Rebassa-Mansergas}}]{McCleery2018}
{McCleery}, J., {Tremblay}, P.-E., {Gentile Fusillo}, N.~P., {et~al.} 2020,
  \bibinfo{title}{{Gaia white dwarfs within 40 pc II: the volume-limited
  Northern hemisphere sample},} \mnras, 499, 1890,
  \dodoi{10.1093/mnras/staa2030}

\bibitem[{S.~A. {Metchev} \& L.~A. {Hillenbrand}(2009){Metchev} \&
  {Hillenbrand}}]{metchev.2009}
{Metchev}, S.~A., \& {Hillenbrand}, L.~A. 2009, \bibinfo{title}{{The
  Palomar/Keck Adaptive Optics Survey of Young Solar Analogs: Evidence for a
  Universal Companion Mass Function},} \apjs, 181, 62,
  \dodoi{10.1088/0067-0049/181/1/62}

\bibitem[{C. {Mordasini} {et~al.}(2012){Mordasini}, {Alibert}, {Klahr}, \&
  {Henning}}]{mordasini.2012}
{Mordasini}, C., {Alibert}, Y., {Klahr}, H., \& {Henning}, T. 2012,
  \bibinfo{title}{{Characterization of exoplanets from their formation. I.
  Models of combined planet formation and evolution},} \aap, 547, A111,
  \dodoi{10.1051/0004-6361/201118457}

\bibitem[{S.~E. {Mullally} {et~al.}(2024){Mullally}, {Debes}, {Cracraft},
  {Mullally}, {Poulsen}, {Albert}, {Thibault}, {Reach}, {Hermes}, {Barclay},
  {Kilic}, \& {Quintana}}]{mullally.2024}
{Mullally}, S.~E., {Debes}, J., {Cracraft}, M., {et~al.} 2024,
  \bibinfo{title}{{JWST Directly Images Giant Planet Candidates Around Two
  Metal-polluted White Dwarf Stars},} \apjl, 962, L32,
  \dodoi{10.3847/2041-8213/ad2348}

\bibitem[{M.~W. {O'Brien} {et~al.}(2023){O'Brien}, {Tremblay}, {Gentile
  Fusillo}, {Hollands}, {G{\"a}nsicke}, {Koester}, {Pelisoli}, {Cukanovaite},
  {Cunningham}, {Doyle}, {Elms}, {Farihi}, {Hermes}, {Holberg}, {Jordan},
  {Klein}, {Kleinman}, {Manser}, {De Martino}, {Marsh}, {McCleery}, {Melis},
  {Nitta}, {Parsons}, {Raddi}, {Rebassa-Mansergas}, {Schreiber}, {Silvotti},
  {Steeghs}, {Toloza}, {Toonen}, {Torres}, {Weinberger}, \&
  {Zuckerman}}]{OBrien2023}
{O'Brien}, M.~W., {Tremblay}, P.~E., {Gentile Fusillo}, N.~P., {et~al.} 2023,
  \bibinfo{title}{{Gaia white dwarfs within 40 pc - III. Spectroscopic
  observations of new candidates in the Southern hemisphere},} \mnras, 518,
  3055, \dodoi{10.1093/mnras/stac3303}

\bibitem[{M.~W. {O'Brien} {et~al.}(2024){O'Brien}, {Tremblay}, {Klein},
  {Koester}, {Melis}, {B{\'e}dard}, {Cukanovaite}, {Cunningham}, {Doyle},
  {G{\"a}nsicke}, {Gentile Fusillo}, {Hollands}, {McCleery}, {Pelisoli},
  {Toonen}, {Weinberger}, \& {Zuckerman}}]{Obrien2024}
{O'Brien}, M.~W., {Tremblay}, P.~E., {Klein}, B.~L., {et~al.} 2024,
  \bibinfo{title}{{The 40 pc sample of white dwarfs from Gaia},} \mnras, 527,
  8687, \dodoi{10.1093/mnras/stad3773}

\bibitem[{M.~W. {Phillips} {et~al.}(2020){Phillips}, {Tremblin}, {Baraffe},
  {Chabrier}, {Allard}, {Spiegelman}, {Goyal}, {Drummond}, \&
  {H{\'e}brard}}]{phillips.2020}
{Phillips}, M.~W., {Tremblin}, P., {Baraffe}, I., {et~al.} 2020,
  \bibinfo{title}{{A new set of atmosphere and evolution models for cool T-Y
  brown dwarfs and giant exoplanets},} \aap, 637, A38,
  \dodoi{10.1051/0004-6361/201937381}

\bibitem[{S. {Poulsen} {et~al.}(2024){Poulsen}, {Debes}, {Cracraft},
  {Mullally}, {Reach}, {Kilic}, {Mullally}, {Albert}, {Thibault}, {Hermes},
  {Barclay}, \& {Quintana}}]{poulsen.2024}
{Poulsen}, S., {Debes}, J., {Cracraft}, M., {et~al.} 2024, \bibinfo{title}{{A
  MIRI Search for Planets and Dust around WD 2149+021},} \aj, 167, 257,
  \dodoi{10.3847/1538-3881/ad374c}

\bibitem[{R.~G. {Probst}(1983){Probst}}]{probst.1983}
{Probst}, R.~G. 1983, \bibinfo{title}{{An infrared search for very low mass
  stars : JHK photometry and results for composite systems.},} \apjs, 53, 335,
  \dodoi{10.1086/190893}

\bibitem[{R.~G. {Probst} \& R.~W. {Oconnell}(1982){Probst} \&
  {Oconnell}}]{probst.1982}
{Probst}, R.~G., \& {Oconnell}, R.~W. 1982, \bibinfo{title}{{The luminosity
  function of very low mass stars.},} \apjl, 252, L69, \dodoi{10.1086/183722}

\bibitem[{S. {Rappaport} {et~al.}(2017){Rappaport}, {Vanderburg}, {Nelson},
  {Gary}, {Kaye}, {Kalomeni}, {Howell}, {Thorstensen}, {Lachapelle}, {Lundy},
  \& {St-Antoine}}]{rappaport.2017}
{Rappaport}, S., {Vanderburg}, A., {Nelson}, L., {et~al.} 2017,
  \bibinfo{title}{{WD 1202-024: the shortest-period pre-cataclysmic variable},}
  \mnras, 471, 948, \dodoi{10.1093/mnras/stx1611}

\bibitem[{M. {Reggiani} {et~al.}(2016){Reggiani}, {Meyer}, {Chauvin}, {Vigan},
  {Quanz}, {Biller}, {Bonavita}, {Desidera}, {Delorme}, {Hagelberg}, {Maire},
  {Boccaletti}, {Beuzit}, {Buenzli}, {Carson}, {Covino}, {Feldt}, {Girard},
  {Gratton}, {Henning}, {Kasper}, {Lagrange}, {Mesa}, {Messina}, {Montagnier},
  {Mordasini}, {Mouillet}, {Schlieder}, {Segransan}, {Thalmann}, \&
  {Zurlo}}]{reggiani.2016}
{Reggiani}, M., {Meyer}, M.~R., {Chauvin}, G., {et~al.} 2016,
  \bibinfo{title}{{The VLT/NaCo large program to probe the occurrence of
  exoplanets and brown dwarfs at wide orbits . III. The frequency of brown
  dwarfs and giant planets as companions to solar-type stars},} \aap, 586,
  A147, \dodoi{10.1051/0004-6361/201525930}

\bibitem[{M. {Rocchetto} {et~al.}(2015){Rocchetto}, {Farihi}, {G{\"a}nsicke},
  \& {Bergfors}}]{rocchetto15}
{Rocchetto}, M., {Farihi}, J., {G{\"a}nsicke}, B.~T., \& {Bergfors}, C. 2015,
  \bibinfo{title}{{The frequency and infrared brightness of circumstellar discs
  at white dwarfs},} \mnras, 449, 574, \dodoi{10.1093/mnras/stv282}

\bibitem[{N. {Skrzypek} {et~al.}(2015){Skrzypek}, {Warren}, {Faherty},
  {Mortlock}, {Burgasser}, \& {Hewett}}]{skrzypek.2015}
{Skrzypek}, N., {Warren}, S.~J., {Faherty}, J.~K., {et~al.} 2015,
  \bibinfo{title}{{Photometric brown-dwarf classification. I. A method to
  identify and accurately classify large samples of brown dwarfs without
  spectroscopy},} \aap, 574, A78, \dodoi{10.1051/0004-6361/201424570}

\bibitem[{P.~R. {Steele} {et~al.}(2011){Steele}, {Burleigh}, {Dobbie},
  {Jameson}, {Barstow}, \& {Satterthwaite}}]{steele.2011}
{Steele}, P.~R., {Burleigh}, M.~R., {Dobbie}, P.~D., {et~al.} 2011,
  \bibinfo{title}{{White dwarfs in the UKIRT Infrared Deep Sky Survey Large
  Area Survey: the substellar companion fraction},} \mnras, 416, 2768,
  \dodoi{10.1111/j.1365-2966.2011.19225.x}

\bibitem[{P.~E. {Tremblay} {et~al.}(2020){Tremblay}, {Hollands}, {Gentile
  Fusillo}, {McCleery}, {Izquierdo}, {G{\"a}nsicke}, {Cukanovaite}, {Koester},
  {Brown}, {Charpinet}, {Cunningham}, {Farihi}, {Giammichele}, {van Grootel},
  {Hermes}, {Hoskin}, {Jordan}, {Kepler}, {Kleinman}, {Manser}, {Marsh}, {de
  Martino}, {Nitta}, {Parsons}, {Pelisoli}, {Raddi}, {Rebassa-Mansergas},
  {Ren}, {Schreiber}, {Silvotti}, {Toloza}, {Toonen}, \& {Torres}}]{Tremblay20}
{Tremblay}, P.~E., {Hollands}, M.~A., {Gentile Fusillo}, N.~P., {et~al.} 2020,
  \bibinfo{title}{{Gaia white dwarfs within 40 pc - I. Spectroscopic
  observations of new candidates},} \mnras, 497, 130,
  \dodoi{10.1093/mnras/staa1892}

\bibitem[{A. {Vigan} {et~al.}(2017){Vigan}, {Bonavita}, {Biller}, {Forgan},
  {Rice}, {Chauvin}, {Desidera}, {Meunier}, {Delorme}, {Schlieder}, {Bonnefoy},
  {Carson}, {Covino}, {Hagelberg}, {Henning}, {Janson}, {Lagrange}, {Quanz},
  {Zurlo}, {Beuzit}, {Boccaletti}, {Buenzli}, {Feldt}, {Girard}, {Gratton},
  {Kasper}, {Le Coroller}, {Mesa}, {Messina}, {Meyer}, {Montagnier},
  {Mordasini}, {Mouillet}, {Moutou}, {Reggiani}, {Segransan}, \&
  {Thalmann}}]{vigan.2017}
{Vigan}, A., {Bonavita}, M., {Biller}, B., {et~al.} 2017, \bibinfo{title}{{The
  VLT/NaCo large program to probe the occurrence of exoplanets and brown dwarfs
  at wide orbits. IV. Gravitational instability rarely forms wide, giant
  planets},} \aap, 603, A3, \dodoi{10.1051/0004-6361/201630133}

\bibitem[{M. {Voyer} {et~al.}(2025){Voyer}, {Changeat}, {Lagage}, {Tremblin},
  {Waters}, {G{\"u}del}, {Henning}, {Absil}, {Barrado}, {Boccaletti},
  {Bouwman}, {Coulais}, {Decin}, {Glauser}, {Pye}, {Glasse}, {Gastaud},
  {Kendrew}, {Patapis}, {Rouan}, {Dishoeck}, {{\"O}stlin}, {Ray}, \&
  {Wright}}]{voyer.2025}
{Voyer}, M., {Changeat}, Q., {Lagage}, P.-O., {et~al.} 2025,
  \bibinfo{title}{{MIRI-LRS Spectrum of a Cold Exoplanet around a White Dwarf:
  Water, Ammonia, and Methane Measurements},} \apjl, 982, L38,
  \dodoi{10.3847/2041-8213/adbd46}

\bibitem[{G. {Yang} {et~al.}(2023){Yang}, {Papovich}, {Bagley}, {Ferguson},
  {Finkelstein}, {Koekemoer}, {P{\'e}rez-Gonz{\'a}lez}, {Arrabal Haro},
  {Bisigello}, {Caputi}, {Cheng}, {Costantin}, {Dickinson}, {Fontana},
  {Gardner}, {Grazian}, {Grogin}, {Harish}, {Holwerda}, {Iani}, {Kartaltepe},
  {Kewley}, {Kirkpatrick}, {Kocevski}, {Kokorev}, {Lotz}, {Lucas},
  {Navarro-Carrera}, {Pentericci}, {Pirzkal}, {Ravindranath}, {Rinaldi},
  {Shen}, {Somerville}, {Trump}, {de la Vega}, {Wilkins}, \& {Yung}}]{yang23}
{Yang}, G., {Papovich}, C., {Bagley}, M.~B., {et~al.} 2023,
  \bibinfo{title}{{CEERS MIRI Imaging: Data Reduction and Quality Assessment},}
  \apjl, 956, L12, \dodoi{10.3847/2041-8213/acfaa0}

\bibitem[{B. {Zuckerman} \& E.~E. {Becklin}(1987){Zuckerman} \&
  {Becklin}}]{zuckerman.1987}
{Zuckerman}, B., \& {Becklin}, E.~E. 1987, \bibinfo{title}{{Excess infrared
  radiation from a white dwarf{\textemdash}an orbiting brown dwarf?},} \nat,
  330, 138, \dodoi{10.1038/330138a0}

\bibitem[{B. {Zuckerman} \& E.~E. {Becklin}(1992){Zuckerman} \&
  {Becklin}}]{zuckerman.1992}
{Zuckerman}, B., \& {Becklin}, E.~E. 1992, \bibinfo{title}{{Companions to White
  Dwarfs: Very Low Mass Stars and the Brown Dwarf Candidate GD 165B},} \apj,
  386, 260, \dodoi{10.1086/171012}

\end{thebibliography}
\bibliographystyle{aasjournalv7}





\end{document}